\documentclass[a4paper,fleqn,usenatbib]{mnras}

\usepackage{newtxtext,newtxmath}

\usepackage[T1]{fontenc}
\usepackage{ae,aecompl}

\usepackage{graphicx}	
\usepackage{amsmath}	
\usepackage{amssymb}	

\usepackage{multirow}
\usepackage{xcolor}
\usepackage{ulem}
\usepackage{cancel}
\usepackage{hyperref}
\usepackage{color}

\title[Local Stability of Galactic Discs in Modified Dynamics]{Local Stability of Galactic Discs in Modified Dynamics}

\author[H. Shenavar, N. Ghafourian]{
Hossein Shenavar$^{1,2}$\thanks{E-mail: h.shenavar@mail.um.ac.ir}
and Neda Ghafourian$^{1,2}$\thanks{E-mail: n.ghafourian@mail.um.ac.ir}
\\
$^{1}$Department of Physics, Ferdowsi University of Mashhad, P.O. Box 1436, Mashhad, Iran\\
$^{2}$ Helmholtz-Institut f{\"u}r Strahlen- und Kernphysik (HISKP), Universit{\"a}t Bonn,  Nussallee 14-16, D-53115 Bonn, Germany
}

\date{Accepted XXX. Received YYY; in original form ZZZ}

\pubyear{2018}

\begin{document}
\label{firstpage}
\pagerange{\pageref{firstpage}--\pageref{lastpage}}
\maketitle

\begin{abstract}
The local stability of stellar and fluid discs, under a new modified dynamical model,  is surveyed by using WKB approximation. The exact form of the modified Toomre criterion is derived for both types of systems and it is shown that the new model is, in all situations, more locally stable than Newtonian model.  In addition, it has been proved that the central surface density of the galaxies plays an important role in the local stability in the sense that LSB galaxies are more stable than HSBs. Furthermore,  the growth rate in the new model is found to be lower than the Newtonian one. We found that, according to this model, the local  instability is related to the ratio of surface density of the disc to a critical surface density $\Sigma^{crit}$.   We provide observational evidence to support this result based on star formation rate in HSBs and LSBs.
\end{abstract}




\section{Introduction}

\label{introduction}
The star formation rate (SFR), which is a key factor in understanding galactic evolution, is usually approximated  by empirical power law of \citet{Schmidt}, which  states that  $\Sigma_{SFR} \propto \Sigma^{n}_{g}$. Here, $\Sigma_{SFR}$ is the SFR per surface density, $\Sigma_{g}$ is the gas surface density and $1 \leq n \leq 3$. \citet{Kennicutt,Kennicutt1}  shows that this law breaks down at densities below a critical threshold value. He also fits the law to a sample of star-forming discs of spiral and starburst galaxies and concludes that $ n = \ 1.4 \pm 0.15 $ for his sample. See also \citet{Elmegreen3} for a review on empirical laws of SFR on galactic scales. On the other hand, gravitational instabilities are usually considered as the source of star formation  and there have been many attempts to relate the instabilities to SFR \citep{Wang,Jog}. Unfortunately though, such attempts are usually faced with difficulties due to little understanding of the star formation at small scales, i.e. the physics of the clouds  \citep{McKee,Wang}. However, one could always hope for an overall correct behaviour derived from gravitational instabilities at larger scales, assuming that the effects of small scale physics could be averaged.

Gravitational instabilities was first studied in detail by \citet{Safronov} who showed that instabilities can occur in a rotating disc of fluid. For the case of stellar discs too, \citet{Toomre64} derives a stability criterion for a razor-thin disc. One key feature  in studying gravitational instabilities is the assumption that wavelength of the perturbation is much smaller than the size of the system, an approximation which is known as WKB or tight-winding approximation \citep{Lin64}. 

There have been many attempts to  make the single fluid or stellar models more realistic. For example, the effect of the thickness of discs is derived by \citet{Vandervoort} while stellar plus gas systems have been considered  by \citet{Kato,Jog3}. Also, a simple stability criterion for stellar plus gas systems has been proposed by  \citet{Wang} and generalized  by \citet{Romeo1}. See also \citet{Elmegreen1} and  \citet{Jog}. Moreover,  \citet{Rafikov} has studied the gravitational stability of systems with several stellar components plus a gas component. All the above mentioned studies have assumed Newtonian force as the governing law of gravitation. Therefore, even with the introduction of a massive halo surrounding a galaxy ( the CDM picture of galactic dynamics), the main features and conclusions of gravitational instabilities will remain intact \citet{Jog2}. 

The standard model of cosmology, i.e. the $ \Lambda CDM$ model,  has been very successful in explaining key properties of galactic and extra galactic scales, though, the most prominent problem of this model is still the detection of the dark matter particle. On the other hand, there have been some attempts to justify galactic dynamics, not by putting more mass to the systems, but by modifying the law of gravity. In this method, the modifications that one introduces to the theory of gravity results in conversion of  the force law  or the inertia term. For the latter, the most famous model is  MOND by \citet{Milgrom1,Milgrom2,Milgrom3} while for the former, among many other theories, one could mention  Scalar-Tensor-Vector theory of \citet{Moffat} (known also as MOG) and the well-known  $f(R)$ theories \citep{Sotiriou}. Among all different proposals to modify gravity, MOND is especially interesting because this model unifies different aspects of galactic dynamics by introducing a single parameter $a_{0}$ which has the dimension of acceleration. See \citet{mond}  for a thorough review on MOND implications, applications and difficulties. 

Theories of modified gravity have to be consistent with the data at different scales. In the case of $f(R)=R^{n}$ theories, in which $R $ is the Ricci scalar, \citet{Sotiriou} have explained that to fit the rotation curve data, the parameter $n$ should depend on the mass of individual galaxies. Thus, one concludes that it is not possible to fit the data for all galaxies with the same choice of $n$. On the other hand,  as \citet{mond} have discussed extensively, although MOND provides a powerful alternative in studying rotationally supported systems, its success in pressure-supported systems is concluded as "less impressive" or even "really problematic" in some cases (e.g. galaxy clusters). Moreover, MOG  explains the rotation curve of spirals \citep{Rahvar} while displays large mass-to-light ratios $M/L$ for dwarf spheroidal (dSph) galaxies  \citep{Haghi}. In addition, concerning the physics of the solar system, some $f(R)$ theories violate the current bounds  on the  perihelion  precession of several planets \citep{Iorio7,Iorio8} while MOND \citep{Milgrom09} and MOG \citep{Moffat08} pass this test.

 Except for consistency with observational data, modified gravities/dynamics with less free parameters are generally  more favoured within the community. In fact,  \citet{Khoury} report that the success of  $\Lambda CDM$ model is based on the fact that only a handful of parameters are required to fit observations.   Also, some modified theories show difficulties in dealing with their mathematical structures. For instance, the non-linearity of MOND  makes it difficult to derive analytical solutions for the model;  though, one could  solve  the equations numerically \citep{mond}. The mathematical simplicity of  models usually leads to easier interpretations of physical consequences; though, there is no guarantee  that the final model must be necessarily simple.

 The problem of local stability of galactic discs in modified theories provides a powerful tool in studying the behaviour of these theories. For example, \citet{Milgrom5} investigates local stabilities of discs governed by  MOND dynamics, \citet{Roshan2,Roshan1} study the effects of MOG while the implications of $f(R)$ have been surveyed  by \citet{Roshan3}.  Also \citet{Roshan4} and \citet{Neda} study the global stability of galactic discs  under MOG from theoretical and numerical point of view respectively.

The aim of this work is to study the theory of local stability, and its observable implications, in a  modified dynamical model. In the next section we will review this modified  model which is based on changing boundary condition of general relativity (GR) from Dirichlet to Neumann \citep{ShenavarI,ShenavarII}. In sections three and four the local stability criterion, growth rate and neutral stability curves are derived for fluid and stellar discs respectively. Most of the physical discussions and observable implications are postponed to Sec. 5 in which we use \citet{Leroy} data  to test the reliability of the model.

\section{\small{Modified Dynamics, A Review}}
\label{MD}
The model which we discuss here, has been derived by assuming a Neumann boundary condition on GR perturbation equations in an expanding universe. See \citet{ShenavarI} for the details. Recently, it has been shown that the surface term of GR action identically vanishes if we assume Neumann boundary condition \citep{Chakraborty,Krishnan}.  \citet{Krishnan2} have also reported an alternative path integral for quantum gravity  using Neumann BC. Furthermore, it could be shown that  a natural BC for gravity in asymptotically AdS spaces is to hold the renormalized boundary stress tensor density fixed instead of fixing the boundary metric \citep{Krishnan3}.

\citet{ShenavarI} imposes Neumann boundary condition on cosmic perturbation equations, i.e. essentially Taylor expansion of Einstein field, and derives a modified Friedmann  and  lensing equations. The reliability of the new lensing equation has been checked by a sample of ten strong lensing systems \citep{ShenavarI}. This new model predicts a  constant acceleration in the equation of motion as $2c_{1}a_{0}$ in which $c_{1}=0.065$ is the Neumann constant and $a_{0}=cH_{0}=6.59 \times 10^{-10} m/s^{2}$. Here $c$ is the speed of light and $H_{0}$ is the Hubble constant. Using this new model one could show that the growth of structures in matter dominated era is more rapid than the standard $\Lambda CDM$ model \citep{ShenavarI}. 

In addition, \citet{ShenavarII} shows that the new term in the equation of motion suggests a small,
though detectable, correction in perihelion  precession of planets. Furthermore, for a system of particles with mass distribution $\rho$ and total mass of M it was shown that the total modified potential is as follows
\begin{equation}   \label{Potential}
\Phi=-G \int \frac{\rho(\vec{x^{\prime}})d^{3}\vec{x^{\prime}}}{| \vec{x^{\prime}} - \vec{x}|}  + \frac{2c_{1}a_{0}}{M}\int \rho(\vec{x^{\prime}})d^{3}\vec{x^{\prime}}|\vec{x^{\prime}} - \vec{x}|
\end{equation} 
in which the first term on the rhs is the gravitational potential of  Newtonian theory while the second one is due to imposing Neumann BC to GR perturbation equations. A potential similar to the second term has been previously named "superpotential" by  \citet{Chandra1,Chandra2,Chandra3}. In fact, these authors  have surveyed many properties of this superpotential in the aforementioned papers. 

The Poisson equation, with which one could start solving many problems in classical mechanics, is modified in this model. In fact, it is easy to derive the next second order Integro-differential equation
\begin{equation}   \label{Second}
\nabla^{2} \Phi = 4 \pi G  \rho + \frac{4c_{1}  a_{0}}{M}  \int \frac{\rho(\vec{x^{\prime}})d^{3}\vec{x^{\prime}}}{| \vec{x^{\prime}} - \vec{x}|}
\end{equation}
as the modified Poisson equation.  However, as it is discussed in \citet{ShenavarIII}, one could see that the next fourth-order Poisson equation 
\begin{equation}    \label{Fourth}
\nabla^{4} \Phi = 4 \pi G \nabla^{2} \rho - \frac{16c_{1} \pi a_{0}}{M} \rho
\end{equation}  
is usually more suitable to use in analysis for the simple fact that it is only differential. In this work too, we will use the fourth-order Poisson equation more often than the second order one.  
 The homogeneous form of  Eq. \eqref{Fourth} is known as biharmonic equation which rises mostly  in the theory of linear elasticity. To solve this equation, one needs four boundary conditions. See \cite{Selvadurai}, chapter 8, for a thorough review on different solutions of the biharmonic equation.  In addition, \cite{biharmonic} present a survey on the existence and uniqueness of solutions of the biharmonic equation.

For a razor-thin disc galaxy with cylindrical  symmetry and radius $R_{d}$, if one assumes an exponential profile for the mass distribution
\begin{equation}       \label{exp}
\Sigma(r) = \Sigma_{0} \exp(-r/R_{d}),
\end{equation}
then it is  possible to derive the rotation curve formula 
 by applying \cite{Casertano1} and \cite{mannheim5} method which is based on direct integration of the potential and using  Bessel function expansion of Green function in cylindrical coordinates. By doing so, one could find \citep{ShenavarII}
\begin{equation}   \label{Rcurve1}
\frac{v^{2} (y)}{GM/R_{d}} =  2y^{2} [I_{0}(y)K_{0}(y)- I_{1}(y)K_{1}(y)]+ \frac{4c_{1}}{\pi}\frac{\Sigma_{\dagger}}{\Sigma_{0}}y^{2}  I_{1}(y)K_{1}(y)  
\end{equation}
in which $y \equiv r/(2R_{d})$ is the scaled radius and  $\Sigma_{\dagger} \equiv a_{0}/G$ is a fundamental surface density in this model which has been previously argued to be an upper limit for the surface density of spirals \citep{Milgrom5}. See \citet{mond} for more discussions on this limit.  The value of $\Sigma_{\dagger}$ is about $9.9 ~kg / m^2$ which is very close to the observational value for central surface brightness of spiral galaxies reported by \cite{Freeman1970}. Now, however, this  is known as a maximum surface density limit above of which the disc galaxies are very rare. See, for instance, \citet{Milgrom5, McGaugh96, mond} and  Sec. \ref{DATA} below for more details. Almost all of our following results show a dependency to this ratio which we call Freeman ratio $\mathcal{R}_{F} \equiv \Sigma_{\dagger} / \Sigma_{0}$. 

Neglecting vertical thickness of the objects, from Eq. \eqref{Rcurve1} one could see that galaxies with  high values of $\mathcal{R}_{F}$ should show rising rotation curves, those with intermediate values of $\mathcal{R}_{F}$ should show a constant circular velocity while objects with the smallest values of  $\mathcal{R}_{F}$ possess declining rotation velocities. Including the thickness of the objects typically decreases the circular velocity. See \citet{ShenavarIII} for the proof of this statement. A similar result is reported by \citet{Casertano} who show that  the rotation curve of low luminosity dwarf galaxies, with  maximum velocity lower than about 100 km/s, are generally rising. Also, the velocities  of intermediate to high luminosity galaxies, with velocities  respectively in the range of $ 100 < v_{max} < 180 km/s$ and $v_{max} > 180 km/s$, are typically flat. But, the rotation velocities of the very highest luminosity galaxies are found generally to be declining from 15 $\%$ for NGC2903 to 30 $\%$ for NGC2683. See \citet{ShenavarII} for the data fitting of 39 LSB galaxies for which it is shown that the rotation curve fittings are generally acceptable.

As we know, the solution to any differential equation, including Einstein field equations, depends on both the functional form of the equation and its boundary condition. The core idea of the present model is to impose Neumann BC, i.e. $\Phi-\Psi=c_1$, on Einstein field equations \citep{ShenavarI} instead of changing the basic  action of GR as it is common in theories of modified gravity.  There are some advantages in this approach, the first of which is that, by doing so the unique Einstein-Hilbert action remains intact. Second,  by assuming Neumann boundary condition the surface term of GR action identically vanishes. This could somehow settle the long-time debate on the surface term of GR  if this assumption leads to a successful cosmological model.  Third,  because the new term in the equation of motion is of the order of the fundamental parameter of MOND, i.e.  $a_{0_{ MOND}}$, the dynamical predictions of the present model too would scale ( in data analysis ) by a similar constant acceleration $a_0$.  Therefore, the success of the present model  in matching observational data could provide a physical interpretation for the clear success of the MOND phenomenology. Fourth, a possible link between local and global physics has been debated for a long time but never been answered thoroughly. See \cite{ShenavarI} for a review on this matter from \cite{Straus} model  to contemporary era. The present model provides another method to build such connection with the possibility to test it with a large amount of observational data at galactic and extragalactic scales. In the following, for instance, we consider the local stability of disc galaxies and we will see that in this model the local stability is correlated with global expansion of the universe through the parameter $a_0$.

 The present model is clearly linear which is a great advantage in analyzing its results. In fact, this property helps to perform the analytical calculations through extending the results of the Newtonian theory in a straightforward manner. For example, we will see  in the following that  the local stability criterion of stellar and fluid systems could be derived. In addition, the present model is built based on only one free  dimensionless parameter, i.e.  $c_1$,  which  could be determined by observations. Thus, this model is quite economical in introducing new parameters.

We will see below that in the present model, the local stability is crucially dependent to $\mathcal{R}_{F}$. Also, it is found that galaxies with low $\mathcal{R}_{F}$ are more unstable than galaxies with high $\mathcal{R}_{F}$. Thus, one should observe that HSB galaxies are more unstable than LSBs. The observational evidence to support this prediction will be presented in Sec. \ref{DATA} which is dedicated to data analysis.

\section{\small{Modified Dispersion Relation for a Fluid Disc}}
\label{DFD}
In a fluid disc, the combined stabilizing effects of pressure and angular momentum are in competition with the  force of gravity which always wants the system to collapse. In a Newtonian model, if the stabilizing effects are dominant, i.e.
\begin{equation}   \label{Qg}
Q_{g} \equiv \frac{\kappa v_{s}}{\pi G \Sigma_{d0}} >1
\end{equation}
then the system would be stable against local collapse \citep{Safronov}. In the last equation, $\kappa$ is the epicyclic frequency which, by using angular velocity $\Omega = v/r$, is defined as follows
\begin{equation}  \label{kappa}
\kappa (r) \equiv \sqrt{ r \frac{d \Omega^{2}}{dr}+4\Omega^{2}}
\end{equation}
 while $v_{s}$ is the sound speed in the fluid  and $ \Sigma_{d0}$ is the disc surface density. One may see \citet{Toomre64} or \citet{BT} page 443 for a physical interpretation of this stability criterion in the context of Newtonian dynamics. 

In the context of dark matter model, \cite{GhoshJog14} have noticed that the square of the total epicyclic frequency  of a disc surrounded by a halo could be written as $\kappa^{2}=\kappa_{disc}^{2}+\kappa_{halo}^{2}$. See Eqs. (17) and (21) of \cite{GhoshJog14} for the exact form of  $\kappa_{halo}$ ( for pseudo-isothermal halo ) and   $\kappa_{disc}$. The presence of the halo would lead to a higher net $\kappa$ and hence a higher $Q$ which results in a more stable system.  In accordance with \cite{GhoshJog14}, we will rewrite the net   epicyclic frequency square as $\kappa^{2}=\kappa_{N}^{2}+\kappa_{c_1}^{2}$ where $\kappa_{N}$ represents the epicyclic frequency due to Newtonian force which  is the same as $\kappa_{disc}$ from \cite{GhoshJog14} while $\kappa_{c_1}^{2}=(2c_1 a_0/R_d) \left[  y(I_0 K_1 - I_1 K_0)+2I_1K_1 \right] $ is resulted from the new term in our modified equation of motion. Due to the existence of the latter term in the present model, the stabilizing effect of  angular momentum is strengthened; thus, one expects a more stable model compared to pure Newtonian one. 

To derive the exact stability criterion in any modified dynamics, one needs the continuity equation  $ \frac{\partial \rho}{\partial t} + \vec{\nabla} .(\rho \vec{v} ) =0$,
the Euler equation $ \frac{\partial \vec{v}}{\partial t} + (\vec{v} . \vec{\nabla})  \vec{v}  = - \frac{ \vec{\nabla} p}{p} -\vec{\nabla} \Phi $ and also the modified Poisson's equation which was introduced in the previous section. We will follow the method by \citet{BT} which is effectively summarized and simplified by \citet{Roshan1}.  The general idea is that one first finds the solution to the Poisson equation for a disc  and then  puts these solutions into the first order approximation of continuity and Euler equations. A barotropic equation of state, i.e. $p=K\Sigma^{\delta}$ in which $K$ and $\delta$ are  real constants, is also assumed here. The system is considered to be an axisymmetric and razor-thin disc. We will use  non-rotating cylindrical  coordinates with $z$ as the rotation axis, while $r$  and $\varphi$ show radial and azimuthal coordinates respectively. Our notation mostly follows \citet{BT}. 

 To linearise the governing equations, we assume that
$ \Sigma_{d} \equiv \Sigma_{d0} + \Sigma_{d1} $, $ v_{r} \equiv v_{r0}+v_{r1}$, $ v_{\varphi} \equiv v_{\varphi 0}+v_{\varphi 1} $, $ \Phi \equiv \Phi_{0}+  \Phi_{1} $   
and the  specific enthalpy $ h    \equiv  h_{0} + h_{1}   $ in which 0 and 1 indices represent  zeroth-order and first order  perturbations respectively. By putting these parameters  into  the continuity equation and Euler equations,  one can rewrite the governing equations  as \citep{BT, Roshan1}:
\begin{equation}   \label{Firstorder}
\begin{split}
&\frac{\partial \Sigma_{d1}}{\partial t} + \frac{1}{r} \frac{\partial (\Sigma_{d0} r v_{r1})}{\partial r} + \Omega \frac{\partial  \Sigma_{d1}}{\partial \phi} +\frac{\Sigma_{d0}}{r} \frac{\partial v_{\varphi 1}}{\partial \varphi}  =0  \\  
&\frac{\partial v_{r1}}{\partial t} + \Omega \frac{\partial v_{r1}}{\partial \varphi} - 2\Omega  v_{\varphi 1} = -\frac{\partial (\Phi_{1} +h_{1})}{\partial r}  \\  
&\frac{\partial v_{\varphi 1} }{\partial t} +\Omega \frac{\partial v_{ \varphi 1}}{\partial \varphi} +\frac{\kappa^{2} v_{r1}}{2 \Omega} = -\frac{1}{r}\frac{\partial (\Phi_{1} +h_{1})}{\partial r}.    
\end{split}
\end{equation}
In addition, by assuming that all perturbations could be approximated locally by a plane wave, i.e. $Q_{1}= Q_{a}e^{i(kr + m\varphi + \omega t)}  $ in which $Q_{1}$ could be any of the perturbations   while $k=2\pi /\lambda$ is the radial wavenumber, and also by presuming WKB approximation $ k \gg m/r$, it is possible to significantly simplify  Eqs. \eqref{Firstorder}. 

 The WKB approximation, also known as tight winding approximation, is assumed to remove the long-range feature of the gravitational force and so it makes the equations local. In this approximation, the radius of the system is much larger than the radial wavelength; thus,  it is possible to omit terms proportional to $1/r$ comparing to the terms proportional to the wavenumber $k$.  Applying this approximation, one can simplify Eqs. \eqref{Firstorder} as
\begin{equation}   \label{Firstorder1}
\begin{split}
&(m\Omega - \omega) \Sigma_{a}+k\Sigma_{d0}v_{ra}=0   \\   
& v_{ra} =\frac{(m\Omega - \omega) k (\Phi_{a}+h_{a})}{\Delta}  \\    
& v_{\varphi a} = \frac{2iB v_{r a}}{\omega - m \Omega}    
\end{split}
\end{equation}
in which
 $ \Delta \equiv \kappa^{2}-(m\Omega - \omega)^{2} $ and $ B(r)  \equiv -\frac{1}{2}\left( \Omega +\frac{d(r\Omega)}{dr}   \right) $
are both functions of radius, known as Oort's parameters, while $h_{a} = v^{2}_{s} \Sigma_{a} / \Sigma_{d0} $ is the amplitude of the specific enthalpy. 

The next step is to find the solution to the first order modified Poisson equation. We use fourth order Poisson equation \eqref{Fourth} instead of the second order Eq. \eqref{Second} essentially because dealing with a pure differential equation is much easier than dealing with an integro-differential equation;  though, in Eq. \eqref{Fourth} one  deals with four boundary conditions instead of two. Here we are seeking the solutions to Eq. \eqref{Fourth} when $\rho = \Sigma_{d1}\delta(z)$ in which $\Sigma_{d1} = \Sigma_{a} e^{i(kr + m\varphi + \omega t)}$.  Assuming without loosing generality that the initial perturbation is in the $x$ direction, one can propose   $\Phi_{1} = \Phi_{a} \exp( i(kx - \omega t) -|\zeta z|) $, in which $\zeta$ is  a constant, to solve the modified Poisson equation. Because there is no mass outside of the plane of the disc, i.e. $ \nabla^{2} \Phi =0 $ when $ z \neq 0$, one can readily show that for these points $\zeta = |k|$ . However, in such system the vertical component of the force is discontinuous in the plane of the disc. Therefore, to find the solution we integrate Eq. \eqref{Fourth} with respect to parameter $z$. The interval of the integration is $z \in (-\zeta, \zeta)$.  See \citet{BT}, chapter two, for a similar treatment of the Newtonian case. The disc is supposed to be razor-thin; therefore  one could find the lhs of Eq. \eqref{Fourth} by taking the limit when $\zeta \to  0$ as 
\begin{equation}   \nonumber
\begin{split}
&\lim_{\zeta \to 0} \int^{\zeta}_{-\zeta} dz \nabla^{4} \Phi_{1} \\    
&= \lim_{\zeta \to 0} \int^{\zeta}_{-\zeta} dz( \frac{\partial^{4} \Phi_{1}}{\partial z^{4}}  +2\frac{\partial^{4} \Phi_{1}}{\partial z^{2} \partial x^{2}}  )\\   
&=  \lim_{\zeta \to 0} (\frac{\partial^{3}}{\partial z^{3}} \Phi_{1} |^{\zeta}_{-\zeta} - 2 k^{2} \frac{\partial}{\partial z} \Phi_{1} |^{\zeta}_{-\zeta})  \\  
& = 2 |\vec{k}|^{3} \Phi_{a} e^{i(kx-\omega t)}
\end{split}
\end{equation}
In the second row  of the above equation we have used the fact that the proposed potential $\Phi_{1}$ is independent of $y$ and continuous with respect to $x$ while in the third row we have differentiated with respect to $x$ and $z$ and also have done the integration.  To obtain the solution to the biharmonic equation, i.e. Eq. (3), we have used the boundary values  of $ \frac{\partial^{3}}{\partial z^{3}} \Phi_{1} $ and  $ \frac{\partial}{\partial z} \Phi_{1} $ at $ \pm \zeta \to 0$ while, because of the symmetry, the second order derivative $ \frac{\partial^{2}}{\partial z^{2}} \Phi_{1} $ is absent from the evaluation on the third row. In addition, the boundary condition on the value of the potential $\Phi$ is  as $ \Phi (0^{+}) =  \Phi(0^{-}) $ which has been implicitly applied before, when we assumed the same amplitude $ \Phi_{a} $ for both sides of the sheet.

The rhs of Eq. \eqref{Fourth} is more straightforward, though one should note that the terms proportional to derivatives of the Dirac delta function converge to zero because $\int f(z) \frac{d \delta (z)}{dz} dz = - \int  \delta (z) \frac{d f(z)}{dz} dz$. The final answer is as follows
\begin{equation}
\Phi_{1} =  \left(  \frac{-2 \pi G \Sigma_{a}}{ |\vec{k}|} + \frac{16 \pi c_{1} a_{0} \Sigma_{a} }{2 M  |\vec{k}|^{3}} \right) e^{i(kx- \omega t ) -|kz|} 
\end{equation}
from which the potential amplitude is read as  $$\Phi_{a} = \frac{-2 \pi G \Sigma_{a}}{ |\vec{k}|}( 1- \frac{4 c_{1}a_{0}}{M  G |\vec{k}|^{2} } )$$.

Now, if we substitute $ \Phi_{a} $ into  the second equation of \eqref{Firstorder1}, we would be able to derive the radial velocity as 
\begin{equation}
v_{ra} = \frac{(m \Omega -\omega)k}{\Delta} \left( \frac{v^{2}_{s}}{\Sigma_{d0}}  -\frac{2 \pi G }{ |\vec{k}|}[ 1- \frac{4 c_{1}a_{0}}{M  G |\vec{k}|^{2} }] \right) \Sigma_{a}
\end{equation}
 By replacing $v_{ra}$ from the last equation into the first equation of \eqref{Firstorder1}  one could find the dispersion relation as 
\begin{equation}  
(m\Omega - \omega)^{2}=\kappa^{2} +k^{2}v^{2}_{s}  -2 \pi G \Sigma_{d0} |\vec{k}|( 1- \frac{4 c_{1}a_{0}}{M  G |\vec{k}|^{2} })
\end{equation}
where in the case of axisymmetric disturbances $m=0$, which is the main focus of the present work, becomes
\begin{equation}  \label{Dispersiong}
\omega^{2}=\kappa^{2} +k^{2}v^{2}_{s}  -2 \pi G \Sigma_{d0} |\vec{k}|( 1- \frac{4 c_{1}a_{0}}{M  G |\vec{k}|^{2} }).
\end{equation}
We should point out that considering $c_{1}=0$ the above  relation reduces to the Newtonian counterpart as we expected. Regarding Eq. \eqref{Dispersiong},  a perturbation with the time dependency proportional to $e^{i \omega t}$ would oscillate forever if $\omega$ is real, or equivalently $ \omega^{2} > 0$, and such  system would be stable. However, if $ \omega^{2} < 0$ then $\omega$ would be a complex number; consequently the perturbation would grow exponentially and thus the system would be unstable. 

The modifications that the present model introduces to the problem of local stability could be well understood by considering the last factor in Eq. \eqref{Dispersiong}, i.e. $A \equiv 1- \frac{4 c_{1}a_{0}}{M  G |\vec{k}|^{2} }$. This factor introduces a boundary wavenumber as $k_{f} \equiv \sqrt{\frac{4 c_{1}a_{0}}{M  G}}$ below of which $A$ is negative and above of that $A$ is positive.  Now, because the first two terms on the rhs of Eq. \eqref{Dispersiong} are positive, if $ k < k_{f}$ then $ \omega^{2} >0 $ and the system is stable against the perturbation with wavenumber $k$. On the other hand, if $ k > k_{f}$ then the last term on the rhs of Eq. \eqref{Dispersiong} would be negative and there would be a competition between stabilizing effects of  angular momentum and  pressure, i.e. $ \kappa^{2} +k^{2}v^{2}_{s} $, and destabilizing effect of gravity. Anyway, in the latter case if gravity won and the system became locally unstable, the unstable modes would grow with a lower rate compared to the pure Newtonian model  because the factor $0< A < 1$ reduces the destabilizing effect of gravity as it is clear from  Eq. \eqref{Dispersiong}.

In what follows in  this section, we will try to demonstrate the different aspects of Eq. \eqref{Dispersiong} in more precise details.  The modified Toomre's criterion, i.e. the local stability criterion, could simply be derived and analysed if we define the dimensionless wavenumber $q$ and the parameter $\beta_{g}$ as
\begin{equation}   \label{gaspar}
\begin{split}
& q \equiv \frac{kv_{s}}{\kappa}  \\   
&\beta_{g} \equiv \frac{v_{s}}{\kappa R_{d}} \\     
\end{split}
\end{equation}
 Using these parameters, we rewrite the dispersion relation \eqref{Dispersiong} as the following equation
\begin{equation}    \label{Dispersiong1}
\frac{\omega^{2}}{\kappa^{2}}= 1+ q^{2} -\frac{2q}{Q_{g}} \left(  1- \frac{2c_{1}}{\pi} \mathcal{R}_{F} \frac{\beta^{2}_{g}}{q^{2}} \right)
\end{equation}
in  which  $ Q_{g} $ is the Newtonian Toomre parameter for a fluid disc which is  defined by Eq. \eqref{Qg}.  It is now easy to see that one could  use the dispersion relation \eqref{Dispersiong1} to rewrite the stability criterion, i.e. $\omega^{2} >1$,  as the following equation
\begin{equation}  \nonumber
Q_{g} > \frac{2q}{1+ q^{2}} \left(   1- \frac{2c_{1}}{\pi} \mathcal{R}_{F} \frac{\beta^{2}_{g}}{q^{2}}      \right) 
\end{equation}
The rhs of the last inequality is dependent to the dimensionless wavenumber $q$. To find a stability criterion independent of $q$ we argue that if $ Q_{g} $ is larger than the maximum value of the rhs of the last equation, then the system is stable for any value of $q$. Thus we rewrite the last equation as
\begin{equation}  \label{Criteriong}
Q_{g} > Max_q \left\lbrace \frac{2q}{1+ q^{2}} \left(   1- \frac{2c_{1}}{\pi} \mathcal{R}_{F} \frac{\beta^{2}_{g}}{q^{2}}      \right) \right\rbrace
\end{equation}
in which $ Max_q  $ represents maximization  with respect to $q$. 
\\
In the case of fluid disc it is possible to find the result of this maximization process analytically. To find the maximum of the rhs of Eq. \eqref{Criteriong}, we will  differentiate the rhs with respect to $q$. By finding the roots of  the result of this differentiation one finds $q_{max}= \pm \frac{\sqrt{1+ 3 b \pm \sqrt{9 b^2+10 b+1}}}{\sqrt{2}} $ where $b= \frac{2c_{1} \beta^{2}_{g}}{\pi} \mathcal{R}_{F} $ has been introduced for the sake of brevity.  Two of the roots are complex numbers for any $b>0$; thus they are dismissed. The other two roots are the same within a minus sign $|q_{max}|=  \frac{\sqrt{1+ 3 b + \sqrt{9 b^2+10 b+1}}}{\sqrt{2}}$. This is in fact the value of the wavenumber $q$  at which the system is closest to being unstable. For this root, the maximum value of the rhs of Eq. \eqref{Criteriong} could be found and thus the stability criterion for fluid disc be rewritten as:
\begin{equation}  \label{newQ}
Q_g > \frac{2 \sqrt{2} \left(b+\sqrt{(b+1) (9 b+1)}+1\right)}{\sqrt{3 b+\sqrt{(b+1) (9 b+1)}+1} \left(3 b+\sqrt{(b+1) (9 b+1)}+3\right)}
\end{equation}
In this way, the stability criterion is written in a way which is manifestly independent of $q$. One could easily check that the rhs of Eq. \eqref{newQ} is a decreasing function of $b$. Also, we notice that for $b=0$, i.e. the Newtonian theory, one finds $q_{max}=1$ and $ Q_g > 1 $ as it is  expected. In addition, 
 through dependency of the parameter $b$ to   $\beta_{g}$ and Freeman ratio $ \mathcal{R}_{F} =\Sigma_{\dagger} / \Sigma_{0}$, we could see that the stability criterion \eqref{newQ} is solely dependent to the location of the point  under consideration and the relative surface density of the galaxy. 
\\
The above results could be obtained from another perspective too. As \cite{Jog3,Jog} have argued, for a system to be in neutral equilibrium, the equations $\omega^{2} (k)=0  $ and $ \dfrac{d \omega^{2} (k)}{dk} =0$ must have a simultaneous real solution for the wave number $k$. By doing so, one could derive the same $q_{max}$ as above and from that the stability criterion Eq. \eqref{newQ} would emerge. We do not perform this calculation here for the sake of brevity.

Now we give an estimation of the magnitude  of $ \beta_{g} $. In the solar neighbourhood, for instance, we have $v_{s} \approx 37 km/s$ and $ \kappa \approx 38 km/s/kpc$ while the radius of the Milky Way is estimated to be about $R_{d} \approx 10 kpc$. Thus, in our neighbourhood we could safely assume that $\beta_{g} \approx 0.1$. For more interior positions, $\kappa$ might rise significantly while the velocity dispersion in the gas $v_{s}$ is still of the same magnitude as before.  Thus $\beta_{g} $ might be much less than 0.1. On the other hand,  in the case of  smaller galaxies we have  $R_{d} \approx 1 kpc$,  but then the ratio of $v_{s}/\kappa$ decreases too as we will see in the  section for data analysis below. Therefore, we consider this parameter to be in the range of $0.01 < \beta_{g} < 0.5$ and we find the maximum value of the rhs of  Eq. \eqref{Criteriong}.
 
 The results of this maximization procedure  are reported in Tab. \ref{table:QqQs} for different values of  $\mathcal{R}_{F}$ and $\beta_{g} $.  For the Newtonian case, i.e. $c_{1}=0$, it is natural to find  the maximum value of the rhs of \eqref{Criteriong}  equal to 1. This is evident from  Eq. \eqref{Criteriong} even without any numerical maximization.  Also we see that when $\beta_{g} $ is relatively small, for example $\beta_{g} =0.01 $, the maximum value is always the same as the Newtonian case. On the other hand, when $\beta_{g} $ increases the maximum value decreases. Thus, regions with larger $\beta_{g} $ are more stable than others.  In addition, low surface brightness galaxies (LSB), i.e. those galaxies with higher ratio of $ \mathcal{R}_{F}$, ought to be  more stable than HSBs because the maximum value of    \eqref{Criteriong} is smaller when $ \mathcal{R}_{F}$ is large. Although to see this vividly, one needs to find places with a higher value of  $\beta_{g} $.

\begin{table*}   
\caption{ Maximum of $Q_{g}$ and $Q_{s}$ for different values of $\beta_{g}$, $\beta_{\star}$ and $ \mathcal{R}_{F} = \Sigma_{\dagger}/\Sigma_{0} $. } 
\centering  
  \begin{tabular}{|c|c|c|c|c|c|c|c|c|c|}   
    \hline 
     \multicolumn{1}{|c|}{}& \multicolumn{3}{c|}{Fluid Systems} &\multicolumn{3}{c|}{Stellar Systems}   \\  \hline
 $ \mathcal{R}_{F} = \Sigma_{\dagger}/\Sigma_{0} $ &  $\beta_{g}=0.01$ & $\beta_{g}=0.1$  & $\beta_{g}=0.5$  &  $\beta_{\star}=0.01$ & $\beta_{\star}=0.1$  & $\beta_{\star}=0.5$   \\
\hline               
$ 0.1 $     & 1.00 & 1.00 & 1.00     &   1.00 & 1.00 & 1.00   \\
$ 1.0 $     & 1.00 & 1.00 & 0.99     &   1.00 & 1.00 & 0.98   \\
$  10.0 $   & 1.00 & 1.00 & 0.88     &   1.00 & 1.00 & 0.87   \\
$  100.0 $  & 1.00 & 0.96 & 0.53     &   1.00 & 0.96 & 0.48   \\
$  200.0 $  & 1.00 & 0.93 & 0.40     &   1.00 & 0.92 & 0.36   \\
\hline 
\end{tabular}
\label{table:QqQs} 
\end{table*}



\begin{figure}
\centering
\includegraphics[height=5cm,width=7cm]{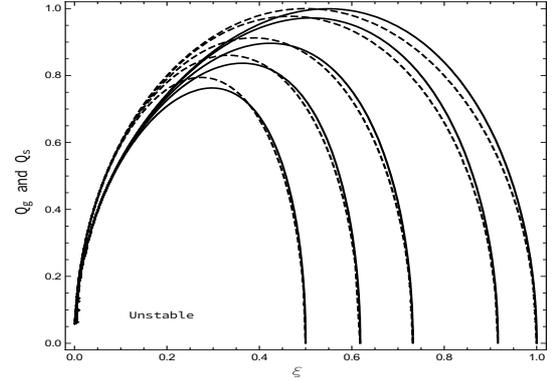}    \\
  \caption{The boundaries of stable and unstable modes for different values of $\gamma$. The neutral boundary for fluid discs $Q_{g}$ is shown by dashed curves while for the stellar discs $Q_{s}$ is shown by solid curves. The top curve is representing the Newtonian case, i.e. $\gamma = c_{1} =0$. The others, from top to bottom, represent $\gamma = "0.1", "0.5", "1.", "2."$ respectively. As it is obvious, the curves for the fluids discs  are very close to the curves of the corresponding stellar discs.   }
  \label{fig:NeutralQgQs}
\end{figure}

The neutral stability curves, i.e. the curves corresponding to $\omega =0$, are interesting too. From Eq. \eqref{Dispersiong1} one could see that defining the dimensionless wavelength $\xi \equiv \lambda / \lambda_{crit}$, in which $ \lambda_{crit} \equiv 4\pi^{2}G\Sigma_{0}/\kappa^{2}$, these curves could be found as 
\begin{equation}     \label{NeutralQgQs}
Q_{g}(\xi )=2\sqrt{\xi(1-\gamma \xi^{2})-\xi^{2}}.
\end{equation}
In this equation $ \gamma \equiv \frac{4 c_{1}a_{0}GM}{(R_{d}\kappa)^{4}} $ is in fact resembling Tully-Fisher relation because it includes on one hand the velocity to the power of four, i.e. $ (R_{d}\kappa)^{4} $, and on the other hand the total mass of the system $M$. However, we do not expect that $\gamma$ be very close to 1 since  for the most parts of the galaxies $R_{d}\kappa$ would not be the same as  flat velocity of the system, for which the Tully-Fisher relation is defined.  Fig. \ref{fig:NeutralQgQs} shows the neutral stability curves versus $\xi$ for different values of $\gamma$. The Newtonian case, which corresponds to $\gamma =c_{1}=0$, is the most unstable case while  by increasing the magnitude of  $\gamma$ more and more regions become completely stable.  

\citet{ShenavarIII} derives the Tully-Fisher relation for the present model   as $V_{f}^{4}=8c_{1}a_{0}\alpha \beta GM$ alongside with three other major scaling relations. Here $\alpha$ and $\beta $ are some dimensionless Virial coefficients while $V_{f}$ represents the flat velocity. See \citet{ShenavarIII} for more details about the general form of the scaling rules in this model  and their test based on data analysis.
 
\begin{figure}
\centering
\includegraphics[height=5cm,width=7cm]{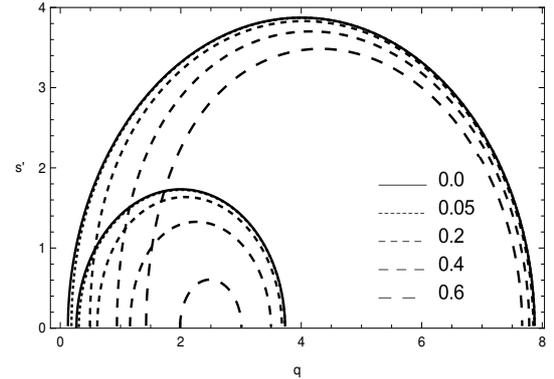}    \\
  \caption{Dimensionless growth rate for a fluid model  with $Q_{g}=0.5 $ ( bottom left curves ) and $Q_{g}=0.25 $ ( the top curves). For all curves we have assumed that $ \mathcal{R}_{F} =100$.  The solid curves show the Newtonian growth rate for which $c_{1} =0.0$.  The other curves represent different values  of $\beta_{g}$ as shown in the plot. The Newtonian theory has a higher growth rate compared to our model. On the other hand, by increasing  $\beta_{g}$, the growth rate decreases. }
  \label{fig:growthrateg}
\end{figure}

Unstable modes, i.e. modes with $\omega^{2} < 0$, show interesting features too. Defining the dimensionless growth rate parameter as $s^{\prime} \equiv \frac{i \omega}{\kappa} $, one could see that Eq. \eqref{Dispersiong1} can be rewritten as
\begin{equation}
s^{\prime 2} = \frac{2|q|}{Q_{g}} \left(    1- \frac{2c_{1}}{\pi}\mathcal{R}_{F} \frac{\beta^{2}_{g}}{ q^{2}}  \right) -(1+q^{2}).
\end{equation}
This quantity  has been plotted in Fig. \ref{fig:growthrateg} as a function of dimensionless wavenumber $q$ for a fixed Freeman ratio $\mathcal{R}_{F} =100$ and different values of $Q_{g} $ and $\beta_{g}$. The growth rate decreases by increasing both $Q_{g} $ and $\beta_{g}$.  The growth rate in the present model is vividly lower  than the Newtonian case. 

The effect of the parameter $\beta_{g}$ on the disc's stability is also evident from Fig. \ref{fig:growthrateg}. It is seen that higher values of $\beta_g = \frac{v_s}{\kappa R_d}$, which happens where the dynamical velocity is comparable to the sound velocity, results in a more stable condition. This is compatible with the observations since, as we will see later, $\kappa$ is lower in the outer parts of the galaxy while  $v_s$ is considered to be almost constant throughout the galaxy; thus, the parameter $\beta_g$ would have a higher value in such regions. Therefore, in the outer parts of the galaxy the stability against local perturbations is stronger, which results in a lower star formation.

\section{\small{Modified Dispersion Relation for a Stellar Disc}}
\label{DSD}
According to \citet{Toomre64}, a stellar disc in Newtonian gravity would be stable if 
\begin{equation}
Q_{s} \equiv \frac{\kappa \sigma_{r}}{3.36 G \Sigma_{d0}} > 1
\end{equation}
in which, again, the stabilizing effects of angular momentum $\kappa$ and velocity dispersion in the radial direction $\sigma_{r}$ are competing with destabilizing force of gravity.  
To find the local stability criterion of stellar discs in a modified gravity/dynamics, one needs the collisionless Boltzmann equation
\begin{equation}
\frac{\partial f}{\partial t} + \vec{v}. \vec{\nabla }f - \vec{\nabla}\Phi . \frac{\partial f}{\partial \vec{v}}=0
\end{equation}
and also the modified Poisson equation \eqref{Second}. The derivation here is similar to the previous section, though, for the reason that the disc is not completely cold, one could not derive  $v_{ra}$ by simply putting $h_{a}=0$ in the second equation of \eqref{Firstorder1}. See \citet{BT} pages 492-495. However, \citet{Toomre64} has shown the existence of a partial cancellation  due to the effects of the spiral potential on $v_{ra}$, i.e. the mean velocity perturbation.   This cancellation reduces the value of $v_{ra}$ by a factor of  $\mathcal{F} \leq 1$, known as the reduction factor, as
\begin{equation} \label{vr}
v_{ra}=\frac{m\Omega -\omega}{\Delta}k\Phi_{a}\mathcal{F}
\end{equation}
For a thorough review on reduction factor and its formal derivation for a razor-thin disc see \citet{BT} appendix K.  The general form of the  factor $\mathcal{F}$ is independent of the form of Poisson equation \citep{Roshan2}. Therefore, it shows the same dependencies on dimensionless frequency and wavenumber and could be written as 
\begin{equation}
\mathcal{F}(s,q^{2})= \frac{1-s^{2}}{\sin \pi s} \int^{\pi}_{0}e^{-q^{2}(1+\cos \tau)} \sin s \tau \sin \tau d\tau
\end{equation}
in which $q \equiv k\sigma_{r}/\kappa$ and $-is^{\prime}=s \equiv (\omega -m\Omega)/\kappa$. 

If we replace $v_{ra}$ from Eq. \eqref{vr} into the first equation of \eqref{Firstorder1}, which by the way remains unchanged, one could derive the following dispersion relation
\begin{equation}
(m \Omega -\omega)^{2} = \kappa^{2}   -2 \pi G \Sigma_{0} |\vec{k}|\left(  1- \frac{4 c_{1}a_{0}}{M  G |\vec{k}|^{2} } \right)  \mathcal{F}(s,q^{2})
\end{equation}
for the stellar disc. Here, we also assume axisymmetric perturbations, for which $m=0$, and so 
\begin{equation}  \label{DispersionS}
\omega^{2} = \kappa^{2}   -2 \pi G \Sigma_{0} |\vec{k}|\left(  1- \frac{4 c_{1}a_{0}}{M  G |\vec{k}|^{2} } \right)  \mathcal{F}(s,q^{2}).
\end{equation}
Recalling the boundary wavenumber $k_{f} = (\frac{4 c_{1}a_{0}}{M  G})^{1/2}$, the rhs of Eq. \eqref{DispersionS} is positive, if $ k < k_{f}$. Thus, in this case $ \omega^{2} >0 $ and the system is stable against local perturbations. On the other hand, if $ k > k_{f}$ then the second term on the rhs of Eq. \eqref{Dispersiong} would be negative and there is a competition between stabilizing effect of  angular momentum and velocity dispersion on one hand, and destabilizing effect of gravity on the other hand. The stabilizing effect of velocity dispersion is included in $\mathcal{F}$.  Anyway, if gravity wins and the system became locally unstable, the unstable modes grow with a lower rate, compared to the pure Newtonian model, because the factor $0< (  1- \frac{4 c_{1}a_{0}}{M  G |\vec{k}|^{2} }) < 1$ decrease the effect of gravity as we see  from  Eq. \eqref{DispersionS}.

Now we derive the modified Toomre's criterion, neutral stability curve and growth rate for the stellar model.  Defining dimensionless parameter $\beta_{\star}$ as
\begin{equation}  \label{stellarpar}
\beta_{\star} \equiv  \sigma_{r}/ (\kappa R_{d})
\end{equation}
one can rewrite the above dispersion relation as follows
\begin{equation}
1=\frac{2\pi |q|}{3.36 Q_{s^{\prime}}} \left( 1- \frac{2c_{1}}{\pi}\mathcal{R}_{F} \frac{\beta^{2}_{\star}}{q^{2}} \right)  \frac{\mathcal{F}(-is^{\prime},q^{2})}{1+s^{\prime 2}}
\end{equation}
The maximum value of the rhs of the above equation occurs at $s^{\prime} =0$. Now, if the rhs is smaller than 1 when $s^{\prime} =0$, then there would be no imaginary solution for $\omega$.  In this case, the local perturbation could not destabilize the disc and the disc would be stable. Thus, for a typical wavenumber $q$ the above equation could be rewritten as
\begin{equation}   \nonumber
Q_{s} > \frac{2\pi |q|}{3.36} \left( 1- \frac{2c_{1}}{\pi}  \mathcal{R}_{F} \frac{\beta^{2}_{\star}}{q^{2}} \right)   \mathcal{F}(0,q^{2})
\end{equation}
To eliminate the dependency on $q$, we notice again that if in the above inequality $ Q_{s} $ was larger than the maximum value of the rhs, then the system would be  stable for any value of $q$. Therefore, we rewrite  the modified Toomre's criterion for the stellar system as
\begin{equation}   \label{Criterions}
Q_{s} > Max_q \left\lbrace \frac{2\pi |q|}{3.36} \left( 1- \frac{2c_{1}}{\pi}  \mathcal{R}_{F} \frac{\beta^{2}_{\star}}{q^{2}} \right)   \mathcal{F}(0,q^{2}) \right\rbrace
\end{equation}
or equivalently, by using modified  Bessel function expression of the reduction factor, i.e. 
$\mathcal{F}(s,\chi) = \frac{1-s^2}{\chi }(1+s^2 e^{-\chi} \Sigma_{n=-\infty}^{\infty}\frac{I_n(\chi)}{n^2-s^2})$, see \cite{BT} page 833, one could derive a more practical form as
\begin{equation}
Q_{s} > Max_q \left\lbrace \frac{2\pi }{3.36 |q|} \left( 1- \frac{2c_{1}}{\pi}  \mathcal{R}_{F} \frac{\beta^{2}_{\star}}{q^{2}} \right)    \left( 1-e^{-q^{2}}I_{0}(q^{2})   \right)  \right\rbrace
\end{equation}
One could easily check that, unlike the fluid disc, the maximization process could not be performed analytically due to complicated dependency of  Bessel function to the wavenumber $q$ through the term $ 1-e^{-q^{2}}I_{0}(q^{2})$. However, the numerical maximization is always possible as we will see in the next section.

\begin{figure}
\centering
\includegraphics[height=6cm,width=8cm]{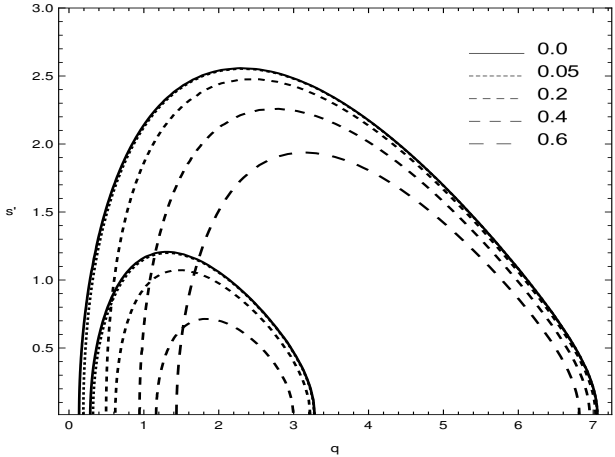}    \\
  \caption{Dimensionless growth rate for a stellar disc   with $Q_{s}=0.5 $ ( bottom left curves)  and $Q_{s}=0.25 $ ( the top curves). In all curves we have considered $\mathcal{R}_{F} =100$.  The solid curves show the Newtonian growth rate for which $c_{1} =0.0$.  The other curves represent different values  of $\beta_{\star}$. The Newtonian theory has a higher growth rate compared to our model. On the other hand, by increasing  $\beta_{\star}$, the growth rate decreases. Also the growth rate in stellar discs are lower than growth rate in fluid models as one could see from comparing this graph with Fig. \ref{fig:growthrateg}}
  \label{fig:growthrates}
\end{figure}
 
The growth rate of the unstable modes of stellar discs, i.e. $\omega^{2} < 0$, is studied by using Eq. \eqref{DispersionS}. The result is plotted for different values of $\beta_{\star}$ as a function of dimensionless wavenumber $q$ in Fig. \ref{fig:growthrates}. It should be mentioned that in this figure, two different values for $Q_s$ is considered and it is obvious that similar to the fluid case, the growth rate of the instability is higher for lower values of Toomre parameter.   Comparing this plot with Fig. \ref{fig:growthrateg} also shows that stellar systems, if unstable, generally collapse with a lower rate than gaseous systems.

As described before, one could also derive the neutral curves, which are the boundary of stable and unstable modes, from the condition $\omega^{2}=0$. Defining for the sake of simplicity a new parameter $f (\xi, Q_{s}(\xi)) \equiv ( \frac{3.36Q_{s}(\xi)}{2 \pi \xi})^{2}$, it is possible to rewrite the dispersion relation \eqref{DispersionS} as
\begin{equation}
\frac{ f }{1-e^{-f}I_{0}(f)} = \frac{(1- \gamma \xi^{2})}{\xi}
\end{equation}
and then plot $Q_{s}$ as a function of $ \xi $. The result is plotted for different values of $\gamma$ by solid lines in Fig.\ref{fig:NeutralQgQs}. The most unstable case is the Newtonian model with $\gamma = c_{1} =0$. In addition,  with increasing $\gamma$ the disc becomes more stable against perturbations with large wavelength $\xi$. In other words, for higher values of  $\gamma$ the total area of instability in $Q-\xi$ plane decreases. Also, the neutral curves of stellar and fluid discs are very similar, though, the peak of the fluid model occurs at a smaller $\xi$ with a slightly larger value.

 The maximum values of Eq. \eqref{Criterions}, for different Freeman ratios $ \mathcal{R}_{F}$ and $\beta_{\star}$, are reported in Tab. \ref{table:QqQs}. In this table we see that when $\beta_{\star} = \beta_{g} $, and for the same Freeman ratio $ \mathcal{R}_{F}$, the maximum values   of the stellar systems are slightly lower than fluid ones. From Fig. \ref{fig:NeutralQgQs} too, one could see that the area of instability region of the fluid system is slightly larger than the stellar one with the same parameter $\gamma$. On the other hand, the Newtonian model  shows a symmetrical behaviour around $\xi =0.5$ for stellar and fluid discs. Thus, one could conclude that in the present model, assuming WKB approximation, the stellar systems are  more stable against local axisymmetric $m = 0$  perturbations  than fluids with same interim parameters.  However, as we will show in the next section, for most galaxies $\beta_{g} $ is an order of magnitude smaller than $\beta_{\star}$; thus, the condition $\beta_{\star} = \beta_{g} $ does not hold in most points of the discs. In addition, to draw a more accurate conclusion about the stability of galactic discs, one needs to treat the galactic discs as a combination of stellar plus fluid systems. We will discuss about this last point in the next section. Moreover, below we will briefly consider the stability of stellar and gaseous discs against non-axisymmetric disturbances $m \neq 0$ independent of $\beta $ parameters. 

It is also possible to use the dispersion relations of fluids and stellar discs to study the behaviour of the dimensionless frequency, i.e.  $\nu \equiv m(\Omega_p -\Omega)/\kappa $ in which $\Omega_p \equiv \omega /m$ is the pattern speed  of an m-armed spiral,  for non-axisymmetric disturbances  $m \neq 0$ . See \cite{BT}, page 497, or \cite{GhoshJog}.  One  expects   the present model to show a different behaviour  at very long ranges, i.e. when $k $ approaches to zero, compared to the Newtonian model    because the gravitating force of the present model is stronger than the Newtonian one. In addition, stellar and fluid systems should have similar trends, because in this regime, the dominant force is gravity while the pressure or velocity dispersion are negligible. From mathematical point of view one could see that for small $k$ we have  $1-\nu^{2}=X(1-\gamma /X^{2})$  in which $X\equiv |k|/k_{crit}$ is the dimensionless wavenumber ( $k_{crit} =2\pi /\lambda_{crit}$ ). Thus, we conclude that in the present model the  dimensionless frequency $\nu$ diverges when $k$ tends to zero. Accidentally, this is the region that the WKB approximation fails; thus, a definitive conclusion needs to go beyond this approximation. 
\\
Moreover, it could be shown that for large wavenumber $k$,  the dimensionless frequency $\nu$ of the stellar systems approaches to $\nu=\pm 1$ (the Lindblad resonances) while  it grows indefinitely for the fluid systems. In this sense, the present model resembles the Newtonian one which  is a predictable behaviour because at large $k$ the dominant factor is no longer self-gravity but pressure - for fluids - or velocity dispersion - for stellar systems - which both remain unchanged in our model. This behaviour shows that at high $k$ fluid systems are more bouncy  compared to stellar systems; thus, former discs  are more stable against non-axisymmetric perturbations  $m \neq 0$. The stability of galactic systems against non-axisymmetric disturbances  needs more careful treatments, though one could see  \citet{Rafikov,GhoshJog} for  thorough investigations on  physical implications of this matter in Newtonian gravity. 

In the context of dark matter paradigm too, \cite{GhoshJog14} have shown that  gravitational instabilities are suppressed  in the case of LSBs dominated by their haloes. They have tested their results for the LSB galaxy UGC 7321, a superthin disc with well-documented observational parameters. The stability of this galaxy against local, linear axisymmetric and non-axisymmetric perturbations has been studied (\cite{GhoshJog14}).  In conclusion, it is proved that in LSBs the star formation and spiral features are mainly suppressed due to  the dynamical effect of the halo  dominating  from inner regions.  The local stability of  a  two-component disc has been shown by \cite{GhoshJog14} too.

\begin{figure*}
\centering
\includegraphics[width=15cm]{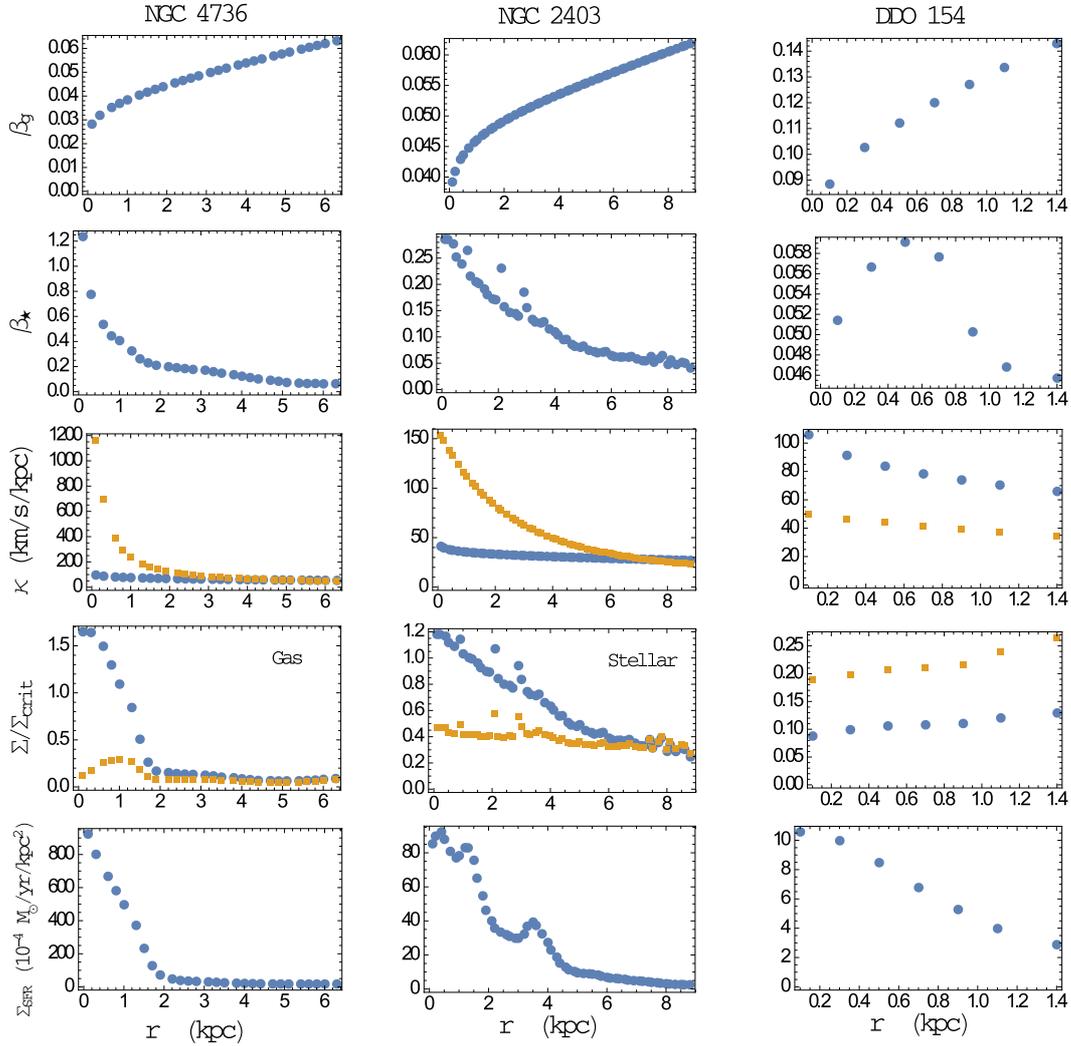}
\caption{ Fundamental parameters of the model. The left column shows the general behaviour of the first sub-sample which is  represented by NGC 4736 galaxy. The middle column displays the general trend of the second sub-sample which is represented by NGC 2403 galaxy. The right column provides a general view on the behaviour of the third sub-sample which is  represented by DDO 154 dwarf irrigular galaxy. Also, the first two rows show dimensionless parameters $\beta_g$ and $\beta_{\star}$ respectively  while the third row  represents epicyclic frequency $\kappa~~(km/s/kpc)$ in which blue points and yellow squares (colour online) display $\kappa$ for our model and  model II ( based on $v_{Boiss}$ ) respectively. The quantity $\Sigma_{SFR}$ - the last row - is mostly correlated with $\Sigma_{g}/ \Sigma^{crit}_{g}$ in the first sub-sample while for the second sub-sample it is closer to the trend of $\Sigma_{\star}/ \Sigma^{crit}_{\star}$. The last sub-sample which consists of three irregular galaxies show no trend. For all galaxies in \citet{Leroy} sample, these parameters  are presented as online plots accompanied by this work. \label{fig:finalplt}}
\end{figure*}

\section{\small{Data Analysis  and Discussion}}
\label{DATA}
In this section, we will use data from \citet{Leroy} to compare the local stability of the present  model with   a second one. Our second model, hereafter named model II, assumes Toomre criterion for which the net epicyclic frequency $\kappa_{net}$ is derived based on  \cite{Boissier} rotation curve formula.  \citet{Leroy} use \citet{Boissier} proposal which describe rotation velocity as a function of radius as
\begin{equation}    \label{Boissier}
v_{Boiss}(r) = v_{flat}( 1 - e^{-r/l_{flat}}).
\end{equation}
In this formula $v_{flat}$ is the velocity at which the rotation curve is flat while $l_{flat}$ is the length scale for which rotation curve begins to be flat. These parameters, for every galaxy in our sample,  are reported in Table. \ref{table:sample}  as derived by \citet{Leroy}. Of course, from the context of the dark matter hypothesis, the flatness of $v_{Boiss}$ at large radii is due to the existence of a dark halo. In the presence of a halo, the general form of the Toomre criterion, i.e. $Q=v_{s}\kappa_{net}/(\pi G \Sigma)$, remains unchanged; though, the $\kappa_{net} $ is derived by combining the contributions from the disc and the halo as explained in Sec. \ref{DFD} above. See Eq. (17) of \citet{Jog2}, for example. However,  \citet{Jog2} derives the contribution of a pseudo-isothermal halo from first principles.

\citet{Leroy}  use HI maps from THINGS ( The HI Nearby Galaxy Survey)  and also derive $H_{2}$ maps. The global properties of this sample  which includes 23 galaxies are introduced in Tab. \ref{table:sample}. We particularly try to find a correlation between the gravitationally unstable positions of the galaxies and their star formation rate surface density $\Sigma_{SFR}$. \citet{Leroy} use  maps of far-ultraviolet (FUV) combined with 24 $\mu m$ maps to find  a tracer sensitive enough to both exposed and dust-embedded star formation. This results in a good measure for $\Sigma_{SFR}$ in units of $10^{-4}M_{\odot}/y/kpc^{2}$. They derive the data related to FUV  from \citet{Paz}. Since very massive  young stars are emitting  their energy mainly through ultraviolet emission, FUV provides a prominent  measure for the rate of star formation.

\begin{table*}
\begin{small}   
\caption{ General properties of our sample including 23 galaxies.  The first sub-sample includes the first 12 galaxies while the next 8 objects represent the second sub-sample. The last three galaxies show our third sub-sample.  Except the four three columns, all other data are reported in Leroy et al. (2008) and they show name, type, radius $R_{d}$ (kpc), stellar mass 
$\log M_{\star}$, HI mass $ \log M_{HI}$, $H_{2}$ mass $\log M_{H_2}$, the radii at which we see a flat rotation velocity $l_{flat}$, the magnitude of the flat rotation velocity $v_{flat}$, a scale length $l_{\star}$ used in fitting $\sigma_{r}$, star formation rate SFR and its exponential scale length $l_{SFR}$.   The last four columns are derived in this work and represent Freeman ratio $ \mathcal{R}_{F} $, the ratio of gaseous mass to the total mass $M_{g}/M_{total}$, the efficiency parameter $\epsilon$ and the  coefficient of determination $R^{2}$. } 
\centering  
  \begin{tabular}{|c|c|c|c|c|c|c|c|c|c|c|c|c|c|c|c|}   
    \hline 
Name & Type &  $R_{d}$ & $\log M_{\star}$ & $ \log M_{HI}$ & $\log M_{H_2}$ & $l_{flat}$ & $v_{flat}$ & $l_{\star}$ & SFR & $l_{SFR}$ &$ \mathcal{R}_{F} $ & $M_{g}/M_{total}$ &    $\epsilon$  & $R^{2}$  \\
     & & (kpc) & $M_{\odot}$ & $M_{\odot}$ & $M_{\odot}$ &  (kpc) & (kpc/s) & (kpc) & $M_{\odot}/y $ & (kpc) & & &  &    \\
\hline               
 \text{NGC7331} & \text{SAb} & 19.6 & 10.9 & 10.1 & 9.7 & 1.3 & 244 & 3.3 & 2.987 & 4.5 & 117. & 0.18 & 0.26 & 0.86 \\
  \text{NGC6946} & \text{SBc} & 9.8 & 10.5 & 9.8 & 9.6 & 1.4 & 186 & 2.5 & 3.239 & 2.7 & 67.7 & 0.25 & 0.20 & 0.98 \\
   \text{NGC5194} & \text{SBc} & 9. & 10.6 & 9.5 & 9.4 & 0.8 & 219 & 2.8 & 3.125 & 2.4 & 52.6 & 0.12 & 0.17 & 0.90 \\
\text{NGC5055} & \text{Sbc} & 17.4 & 10.8 & 10.1 & 9.7 & 0.7 & 192 & 3.2 & 2.123 & 3.1 & 110.9 & 0.22 & 0.15 & 0.93 \\
\text{NGC4736} & \text{Sab} & 5.3 & 10.3 & 8.7 & 8.6 & 0.2 & 156 & 1.1 & 0.481 & 0.9 & 39.8 & 0.04 &  0.51 & 0.99 \\
 \text{NGC3627} & \text{SBb} & 13.9 & 10.6 & 9. & 9.1 & 1.2 & 192 & 2.8 & 2.217 & 1.9 & 135.8 & 0.05  & 0.07 & 0.69\\
 \text{NGC3521} & \text{SBbc} & 12.9 & 10.7 & 10. & 9.6 & 1.4 & 227 & 2.9 & 2.104 & 3.1 & 76.7 & 0.22  & 0.21 & 0.88 \\ 
 \text{NGC3351} & \text{SBb} & 10.6 & 10.4 & 9.2 & 9. & 0.7 & 196 & 2.2 & 0.94 & 1.8 & 119.9 & 0.09   & 0.95 & 0.98\\ 
  \text{NGC3198} & \text{SBc} & 13. & 10.1 & 10.1 & 8.8 & 2.8 & 150 & 3.2 & 0.931 & 3.4 & 193.6 & 0.51 &  0.93 & 0.43 \\
   \text{NGC3184} & \text{SBc} & 11.9 & 10.3 & 9.6 & 9.2 & 2.8 & 210 & 2.4 & 0.901 & 2.8 & 164.1 & 0.22 &  0.46 & 0.77\\
 \text{IC2574} & \text{Irr} & 7.5 & 8.7 & 9.3 & 7.9 & 12.9 & 134 & 2.1 & 0.07 & 4.8 & 645.6 & 0.81  &  3.07   & 0.81\\
 \text{NGC628} & \text{Sc} & 10.4 & 10.1 & 9.7 & 9. & 0.8 & 217 & 2.3 & 0.807 & 2.4 & 171.9 & 0.32  &  0.38 & 0.84\\
 \hline
 \text{NGC925} & \text{SBcd} & 14.2 & 9.9 & 9.8 & 8.4 & 6.5 & 136 & 4.1 & 0.561 & 4.1 & 411. & 0.45  &  1.18 & 0.53 \\
 \text{NGC2403} & \text{SBc} & 7.3 & 9.7 & 9.5 & 7.3 & 1.7 & 134 & 1.6 & 0.382 & 2. & 192.3 & 0.39   &  0.36 & 0.84 \\
 \text{NGC2841} & \text{Sb} & 14.2 & 10.8 & 10.1 & 8.5 & 0.6 & 302 & 4. & 0.741 & 5.3 & 78.4 & 0.17   &  0.37 & 0.78 \\
 \text{NGC2976} & \text{Sc} & 3.8 & 9.1 & 8.3 & 7.8 & 1.2 & 92 & 0.9 & 0.087 & 0.8 & 280.6 & 0.17  &  0.36 & 0.79 \\
 \text{NGC3077} & \text{Sd} & 3. & 9.3 & 9.1 & 6.5 & - & - & 0.7 & 0.086 & 0.3 & 81.7 & 0.39   &  0.21 & 0.4\\
 \text{NGC4214} & \text{Irr} & 2.9 & 8.8 & 8.7 & 7. & 0.9 & 57 & 0.7 & 0.107 & 0.5 & 217.7 & 0.45   &  0.37 & 0.4 \\
 \text{NGC4449} & \text{Irr} & 2.8 & 9.3 & 9.2 & 6.9 & - & - & 0.9 & 0.371 & 0.8 & 64.6 & 0.44   &  0.43 & 0.73 \\
 \text{NGC7793} & \text{Scd} & 6. & 9.5 & 9.1 & 0 & 1.5 & 115 & 1.3 & 0.235 & 1.3 & 240.7 & 0.28  &  0.29   & 0.80\\
 \hline
\text{DDO154} & \text{Irr} & 1.2 & 7.1 & 8.7 & 6.8 & 2. & 50 & 0.8 & 0.005 & 1. & 81.9 & 0.98 & - & - \\
 \text{HoI} & \text{Irr} & 1.8 & 7.4 & 8.3 & 7.2 & 0.4 & 53 & 0.8 & 0.009 & 1.2 & 398.3 & 0.9  & - & - \\
 \text{HoII} & \text{Irr} & 3.7 & 8.3 & 8.9 & 7.6 & 0.6 & 36 & 1.2 & 0.048 & 1.3 & 391.5 & 0.81 & - & - \\
 \hline 
\hline 
\end{tabular}
\label{table:sample} 
\end{small}
\end{table*}

\textit{\textbf{ Deriving epicyclic frequency $\kappa$ for the two models:}} To study the local stability of galaxies in any specific point,  one needs the value of epicycle frequency $\kappa$. In the present model, knowing that $\Omega \equiv v/r$, one could find the square of the angular velocity $\Omega$ from Eq. \eqref{Rcurve1} as
\begin{equation}  
 \frac{\Omega^{2} (y) }{GM/(4R^{3}_{d})} =  2 [I_{0}(y)K_{0}(y)- I_{1}(y)K_{1}(y)]  
       + \frac{4c_{1}}{\pi} \mathcal{R}_{F}  I_{1}(y)K_{1}(y).  
\end{equation}
 Then, it is possible to derive the value of $\kappa$ at different radii in any galaxy as we discussed after Eq. \eqref{kappa}. We have plotted $\kappa$ as a function of radius for three galaxies, i.e. NGC 4736, NGC 2403 and DDO 154,  in Fig. \ref{fig:finalplt}. For the rest of the sample, the behaviour of $\kappa$ is also derived and presented as online figures.  By using $ v_{Boiss} $ from Eq. \eqref{Boissier}  one can also find  epicycle frequency for   model II  as $\kappa_{Boiss}(r) = 1.4\frac{v_{Boiss}(r)}{r} \sqrt{1+a}$ in which $a=\frac{r}{v_{Boiss}} \frac{dv_{Boiss}}{dr} $. We have plotted $\kappa_{Boiss}$ in Fig. \ref{fig:finalplt} alongside with $\kappa_{MOD}$.   The results for the present model are shown  by blue points while the behaviour of  model II is displayed by yellow squares (color online). As it is clear, our model predicts a relatively constant $\kappa$ compared to model II. At the centre, however, the magnitude of $\kappa$ rises. We will see in the following that the  epicyclic frequency is, in fact, the dominant factor in determining the general behaviour of $\Sigma_{SFR}$. Also, the difference between our model and  model II is mostly due to the difference in $\kappa$ in these two models.

\textit{\textbf{  The effects of $\beta_{g}$ and $\beta_{\star} $ on stability:}} To find a relation between instabilities and $\Sigma_{SFR}$, we point out that the stability of the present model is dependent to   $\beta_{g}$ and $\beta_{\star} $ for gaseous and stellar systems respectively and also Freeman ratio $\mathcal{R}_{F}$.  To find $\beta_{g}$ from Eq. \eqref{gaspar} one needs the sound speed $v_{s}$ of the gas which we assume to be equal to the gas velocity dispersion. \citet{Leroy} report that for their sample, the gas velocity dispersion is almost constant and about $11 \pm 3 ~ km/s $. The uncertainty in this  quantity introduces about $27 \% $ error into our results. In addition, for the  radius of the disc $R_{d}$, we use the optical radius which is reported in Tab. \ref{table:sample} ( also derived from Tab. 4 of \citet{Leroy}). It should be noted that this radius, being  defined by an optical isophote,  measures the radius of stellar light but  not the radius of gas component. The gaseous component is usually distributed to larger radii, so the actual radius of the galaxies are considered to be larger than optical radius. The difference, of course, depends on the ratio of $M_{gas}/M_{\star}$ and the distribution of gas which could be very different from object to object. The point is that this issue could introduce another source of uncertainties into our estimations. Due to these sources of error, one could only hope for an overall agreement between the theory and observation.

Knowing the behaviour of $\kappa$,  it is easy to see that at large radii $r$, the $\beta_{g}$ parameter increases as  $ \beta_{g} \propto \sqrt{r}$ for the present model while for small radii it converges to zero. The values of $\beta_{g}$ are shown in Fig. \ref{fig:finalplt} for three chosen galaxies.  At small radii $\beta_{g}$ is  small, therefore,  we expect more instabilities in  the gas component in such regions compared to large radii. In other words, at small $r$ one needs a larger Toomre value to provide the stability as we explained in Tab. \ref{table:QqQs}. On the other hand, at large radii the parameter $\beta_{g}$ rises as square root of radius; thus gas becomes more stable.

Now we estimate  the parameter $\beta_{\star}$ from Eq. \eqref{stellarpar}. To do so, we need the value of  radial stellar velocity dispersion  $\sigma_{r}$ at different radii. \citet{Leroy} have reported that the vertical  stellar velocity dispersion  $\sigma_{z}$ and the radial one are related through the equation $\sigma_{z}=0.6 \sigma_{r}$. From this observation, and the reported value for $\sigma_{z}$ in \citet{Leroy}, one could see that
\begin{equation}
\sigma_{r} (r) = 1.55 \sqrt{Gl_{\star} \Sigma_{\star} (r)}
\end{equation}
where $ l_{\star} $ is another observational parameter with the dimension of length  and $ \Sigma_{\star} $ is the surface density of stars which decreases exponentially toward outer parts of the objects. Both parameters are reported by \citet{Leroy}.  The other two parameters which we need to estimate  $\beta_{\star}$ at different points, i.e. $\kappa$ and $R_{d}$, were discussed above. Putting all these parameters together, one expects that $\beta_{\star}$  tends to zero at very small ( almost zero ) radii, then rises very steeply due to the exponential behaviour of stellar surface density until a maximum value, then again decreases at large radii.  The result for  $ \beta_{\star}$  are shown in Fig. \ref{fig:finalplt}.  For the rest of the galaxies of Tab. \ref{table:sample}, the parameters $\beta_g$ and $\beta_{\star}$ are presented as online plots. In general due to the lack of the data at very small radii, we do not see the tendency toward zero near the centres of the objects. However, the curves show a maximum value and a general exponential decay.  Thus, for the stellar disc the rhs of Eq. \eqref{Criterions} is smaller toward the center. See also Tab. \ref{table:QqQs}.

\textit{\textbf{Rewriting stability criterion to include model II:}}  
To compare the results of the present model with the results of model II, we first need to introduce a new measure for stability/instability instead of  $\beta$  because model II is not dependent to this parameter. To do so, we rewrite the stability criterion as
\begin{equation} \label{StableSigma}
\Sigma(r) < \Sigma^{crit} (r)
\end{equation}
in which the critical surface density $\Sigma^{crit} (r)$ for our theory is as
\begin{equation}
\Sigma^{crit}_{MOD} \equiv \left(  \frac{\Sigma^{crit}_{N}}{H} \right)_{g~or~\star}
\end{equation}
while the Newtonian critical surface density for gaseous and stellar components are respectively as
$\Sigma^{crit}_{N ~g} \equiv \kappa v_{s}/(\pi G)$
and
$\Sigma^{crit}_{N~ \star} \equiv \kappa \sigma_{r}/(3.36 G)$. The function $H_{g}$ could be found from Eq. \eqref{Criteriong} as
\begin{equation}
H_{g} ( \mathcal{R}_{f} , \beta^{2}_{g}) \equiv Max_q \left\lbrace   \frac{2 q}{1+ q^{2}} \left(   1-\mathcal{R}_{F} \frac{2c_{1}}{\pi} \frac{\beta^{2}_{g}}{q^{2}} \right) \right\rbrace  <1
\end{equation}
for the gaseous component while from Eq. \eqref{Criterions} we have
\begin{equation}
H_{\star}( \mathcal{R}_{f}, \beta^{2}_{\star})  \equiv Max_q \left\lbrace \frac{2\pi |q|}{3.36} \left( 1- \mathcal{R}_{F} \frac{2c_{1}}{\pi} \frac{\beta^{2}_{\star}}{q^{2}} \right)   \mathcal{F}(0,q^{2}) \right\rbrace <1 
\end{equation}
for the stellar disc.  The criterion \eqref{StableSigma} is a well motivated stability criterion because it states that wherever the local surface density $\Sigma(r)$ is larger than the critical surface density $\Sigma^{crit} (r)$ there would be a chance of instability. The ratio $ \Sigma^{crit} / \Sigma$ is in fact Toomre's parameter for gaseous and stellar discs.   $\Sigma^{crit}$ combines the effects of $\beta$ parameters, Freeman ratio $\mathcal{R}_{F}$ and the epicyclic frequency $\kappa$.

 The $H$ functions does not dependent to dimensionless wavenumber $q$ due to maximization.   The possible values of $q$ are bounded from below by $ \sqrt{\mathcal{R}_{F} \frac{2c_{1}}{\pi}} \beta $ because $H$ could not be negative in stable modes. The maximization results in an absolute minimum   value of critical surface density $ \Sigma^{crit} (r) $ below of which the gas is relatively safe from collapse. We say "relatively safe" because, after all, we work in WKB regime which is consistent with relatively large wavenumber $q$, i.e. small wavelength. Any perturbation with large wavelength $\lambda$, large compared to the size of the system, could still make system unstable and force the local collapse. This could happen when nearby galaxies interact.  While the values of $\beta$ parameters and Freeman ratio $\mathcal{R}_{F}$ are important in determining the $H$ function, we will see in the following that $\kappa$ parameter is the dominant factor in determining  $ \Sigma / \Sigma^{crit}  $ and thus the stability of the systems.

\textit{\textbf{  A relation between $\Sigma_{SFR}$ and surface density ratio:}}
The physics of Eq. \eqref{StableSigma} is simple. In any model, if this criterion is violated, there is local instability and so one could observe collapse of the matter which could lead to star formation. Of course, as emphasized before, this statement is true in the domain of the validity of the WKB approximation.  To find a relation between star formation rate surface density $\Sigma_{SFR}$ and local instabilities, we have plotted   for three different galaxies  in  Fig. \ref{fig:finalplt} the ratio of stellar surface density to stellar critical surface density, i.e.  $\Sigma_{\star}/\Sigma^{crit}_{\star}$, and the similar ratio for gas component $\Sigma_{g}/\Sigma^{crit}_{g}$. The result for other objects in \citet{Leroy} sample is presented as online figures accompanying this work. The blue points show the results predicted by the present model while the yellow squares are based on  model II. In Fig. \ref{fig:finalplt}, the last row shows   $\Sigma_{SFR}$ in units of $10^{-4}M_{\odot}/year/kpc^{2}$ as reported by \citet{Leroy}.

 $\Sigma_{SFR}$ is a particularly interesting measure to be compared with $\Sigma / \Sigma^{crit}$ because both parameters are defined locally. In other words, theses are not global quantities  related to the objects as general, e.g. total SFR; thus, point to point comparison  is possible. In our following discussions $\Sigma_{g}$ represents the total gaseous mass surface density, i.e. mass surface density of HI  plus that of $H_{2}$,  reported by \citet{Leroy}.

We have divided THINGS catalogue to three distinctive groups based on the correlation between $\Sigma_{SFR}$ and $\Sigma_{g}/\Sigma^{crit}_{g}$ or $\Sigma_{\star}/\Sigma^{crit}$.  The majority of the sample show   an overall correlation between $\Sigma_{SFR}$ and the gas ratio $\Sigma_{g}/\Sigma^{crit}_{g}$  predicted by our theory. This includes 12 galaxies which are represented by NGC 4736 on the left column of Fig. \ref{fig:finalplt}.  We propose that in this sub-sample there is a linear correlation between $\Sigma_{SFR}$ and the gas surface ratio as  $ \Sigma_{SFR} \propto \Sigma_{g}/\Sigma^{crit}_{g}$.   

 The correlation between $\Sigma_{SFR}$, i.e. the fifth row in Fig. \ref{fig:finalplt}, and $\Sigma_{g}/\Sigma^{crit}_{g}$, i.e. the fourth row in Fig. \ref{fig:finalplt},  derived from  model II ( yellow squares ) is  poor or even negligible in the first sub-sample. The problem is that the predicted value for $\Sigma_{g}/\Sigma^{crit}_{g}$,  derived based on $v_{Boiss}$,  is almost constant as a function of radius.   In other words, the critical surface density in  model II is almost the same for different positions while we see that the $\Sigma_{SFR}$ increases sharply toward the center and it decays toward outer parts of the galaxies.  For other objects in this sub-sample, there are even cases where we see a high growth of $\Sigma_{SFR}$ toward center while model II predicts a decrease in $\Sigma_{g}/\Sigma^{crit}_{g}$. Most notably, we could mention NGC 4736, NGC 5055, NGC 628 for this behaviour. See the online plots. However, in some other cases, the general trends of $\Sigma_{g}/\Sigma^{crit}_{g}$ in model II and the present one  are similar while the latter predicts a higher ratio at the center. 

The behaviour of the irregular galaxy IC 2574 is particularly different in this sub-sample because in this case the   model II predicts a larger ratio of $\Sigma_{g}/\Sigma^{crit}_{g}$ compared to the present model. However, $\Sigma_{SFR}$ tracks both models with good concordance.

In the second group of galaxies, which contains eight objects, we observe that $\Sigma_{SFR}$ follows the trend of  $\Sigma_{\star}/\Sigma^{crit}_{\star}$ instead of  $\Sigma_{g}/\Sigma^{crit}_{g}$.  This sub-sample is represented by NGC 2403 galaxy for which we have displayed the relevant parameters in the middle column of Fig. \ref{fig:finalplt}.  The stellar ratio $\Sigma_{\star}/\Sigma^{crit}_{\star}$ in this subgroup is sensibly larger than the the gaseous one $\Sigma_{g}/\Sigma^{crit}_{g}$; thus we assume that the gas instability in these systems are mostly driven by stellar instabilities.  

 We should point out that for two of the  objects in this subgroup, namely NGC 3077 and NGC 4449, \citet{Leroy} provide no $l_{flat}$ and $v_{flat}$ because of their complex velocity fields.  As a result, the  prediction of model II for $\Sigma / \Sigma^{crit}$ was undefined here. Also, for the galaxy NGC 2841 there were no  data of the gas surface density in the inner parts up to $3.1 kpc $. We put zero for the magnitude of the surface density of $HI$ and $H_{2}$; thus, in online figures we have $\Sigma_{g}/\Sigma^{crit}_{g}=0$ for $r< 3.1~kpc $ for this object. Anyway, the star formation rate surface density still shows a good agreement with $\Sigma_{\star}/\Sigma^{crit}_{\star}$  in the absence of data for $\Sigma_{g}$ in this case. For this group we observe an approximate correlation, i.e. weaker than the previous sub-sample, of the form $ \Sigma_{SFR} \propto \Sigma_{\star}/\Sigma^{crit}_{\star}$. In the second group too, the correlation between $\Sigma_{SFR}$ and $\Sigma_{\star}/\Sigma^{crit}_{\star}$ derived from our model is better than the prediction of model II.

Finally, there is a third group, containing three irregulars, which shows no correlation between $\Sigma_{SFR}$ and $\Sigma_{g}/\Sigma^{crit}_{g}$ or $\Sigma_{\star}/\Sigma^{crit}_{\star}$. In fact, as it is shown for DDO 154, the parameters $\Sigma_{g}/\Sigma^{crit}_{g}$ and $\Sigma_{\star}/\Sigma^{crit}_{\star}$ are very low compared to other objects. This results in low SFR. However, the reason that these ratios are very low is that their $\Sigma^{crit} $ is very high due to their very high epicyclic frequency $\kappa$. In fact, one could easily show that the amplitude of $\kappa_{c_1}$, i.e. $2c_1 a_0 /R_d$, is at least five times higher than  the amplitude of Newtonian epicyclic frequency $\kappa_{N}$, i.e. $GM/(2 R^{3}_d)$. The radius of the gas component is reported as $R_d = 4.96~ kpc $ by \citet{Sparc}.   Therefore, we conclude that although this system is very gas rich, star formation  has not yet been started in this galaxy because of the high value of the term $\kappa_{c_1}$ which leads to high  $\Sigma^{crit}$ and thus a small $\Sigma/\Sigma^{crit}$. In the context of dark matter model too, \citet{Carignan} have  reasoned that the low SFR of DDO 154 might be due to presence of the potential well of a dark halo.

As it was mentioned before, for the first two groups of galaxies, which include 20 out of 23 objects, a good law for $\Sigma_{SFR}$ could be stated as $\Sigma_{SFR} = A \frac{\Sigma}{\Sigma^{crit}}$. In this equation, the  surface density ratio  $\frac{\Sigma}{\Sigma^{crit}} $ is equal to $\Sigma_{g}/\Sigma^{crit}_{g}$ for the first subgroup while it equals $\Sigma_{\star}/\Sigma^{crit}_{\star}$ for the second one.  The amplitude of $\Sigma_{SFR}$, i.e. the parameter $A$ which is measured in $M_{\odot}/year/kpc^{2}$, could be found by integrating the observational data for $\Sigma_{SFR}$ and also $ \frac{\Sigma}{\Sigma^{crit}} $ over all radii as $A = \frac{\int~\Sigma_{SFR}~dr}{\int ~\Sigma / \Sigma^{crit}~dr }  $. Then, as our final result, we derived the value of $A$ for all galaxies and find 
\begin{equation} \label{lSFR}
\Sigma_{SFR} = 10^{4} \epsilon \frac{SFR}{2 \pi l^{2}_{SFR}}   \frac{\Sigma}{\Sigma^{crit}}
\end{equation}
in which $SFR$ is the total star formation rate (in $M_{\odot}/year$ ) while  $l_{SFR} =  (1 \pm 0.2)l_{\star}$ (in $kpc$) is the SFR exponential scale lengths. Both quantities are determined for different galaxies by \citet{Leroy} and also reported here in Tab. \ref{table:sample}. The factor $10^{4}$ appears to balance the units of the two sides of Eq. \eqref{lSFR}.  The dimensionless parameter $\epsilon $  is derived from a linear   fit model of Eq.  \eqref{lSFR} and  reported in Tab. \ref{table:sample} for all objects. This parameter is found to be close to unity in all cases and might be related to the small scale efficiency of star formation for different objects. Because the small scale details of star formation, i.e. molecular clouds scale and clumps within them, is yet not understood well  \citep{McKee,Wang}, the possible physical meaning of parameter $\epsilon $ needs a more careful and independent study. We should point out that one might interpret the parameter $\epsilon $ as a  constant of proportionality due to uncertainties in the magnitude of SFR and $l_{SFR}$. This might be the case, however, theses uncertainties could at most be about $50~ \%$  while the value of $\epsilon$ changes from $0.07$ up to $3.07$. Therefore, the uncertainties in SFR and $l_{SFR}$ could not be held solely responsible for the existence of the parameter $\epsilon $.

Assuming that SFR  is proportional to the total gas surface density, then   Eq. \eqref{lSFR} suggests that the power $n$ in Schmidt-Kennicutt law is equal to 2 in the first group while it is equal to 1 in the second one. Therefore, for instance, if a sample  includes an equal share  of galaxies with stellar discs and gaseous discs dominating the local instabilities, then one expects an average of $n \approx 1.5$ as \citet{Kennicutt1} proposes. However, if the share of the first sub-sample is significantly larger than the second one, then we expect $n$ to be larger than 1.5. In the same way, $n < 1.5$ if the share of the second group is dominant.  \citet{Elmegreen3} proposes that when $\Sigma_{HI} > \Sigma_{H_2} $ then $\Sigma_{SFR} \propto \Sigma_{g}\times \Sigma_{\star}$; thus, the multiplication of the gaseous and stellar surface density as the governing law of star formation rate is observed before.

In comparison with other galaxies, it could be seen that for three irregulars of the third sub-sample, although  there is no trend of the form of Eq. \eqref{lSFR}, the range of $\Sigma_{SFR}$ is compatible with $\Sigma/\Sigma^{crit}_{\star~or~g}$ . Therefore, the low amount of SFR in these gas-rich low surface density galaxies could be explained without the need of a DM halo. This is because of the high value of $\kappa$ as explained above for the case DDO 154.

In Tab. \ref{table:sample} we have reported the coefficients of determination $R^2$ of the linear model Eq. \eqref{lSFR} for all objects. It could be seen that for the first sub-sample the highest value of  $R^2$ is equal to 0.99 while ten - out of twelve - objects show an $R^2$ larger than 0.7. The lowest $R^2$ in this sub-sample is about 0.43.  The second sub-sample, on the other hand, shows smaller $R^2$ with the highest value of 0.84 and lowest value of 0.4. Five objects, out of eight, show $R^2$ larger that 0.7. In conclusion,  Eq. \eqref{lSFR} shows a general qualitative agreement, though, its behaviour is far from ideal. The problematic regions of the fit are mostly found in outer parts of the galaxy as one could see from our online plots.  It should be noted that in this work we have only considered the the role of gravitational instabilities in the star formation. In fact, as \citet{McKee} have explained in details, the star formation might also be affected by turbulence, magnetic fields and the environment. Including these parameters is beyond the scope of the present work. However, as we discuss briefly below, the problematic behaviour at large radii could be improved by building a more accurate gravitational model.

\textit{\textbf{ Problem at large radii:}}
Clearly, the objects of Fig. \ref{fig:finalplt}   show a damping  $\Sigma_{SFR}$  toward outer radii; though, this is also correct for most objects in Tab. \ref{table:sample}.  The damping of  Eq. \eqref{lSFR} on the other hand,  is not quite enough as one expects from a completely reliable model, although it certainly decreases at larger radii.  This defect, could be related to two major assumptions of our calculations. The first one is that we haven't considered the thickness of the objects  in our formulation and the second is the single-component assumption of the present work.  The effects of the thickness  has been considered before, in the context of Newtonian gravity, from different points of view. \citet{Elmegreen2}, for example, have studied the effect of gravitational instabilities on the star formation for dwarf irregular galaxies and also the outer regions of the spirals. They claim that since the discs get thick in the outer parts, the conventional 2D gravitational instability using Toomre's criterion is not enough to explain the observed star formation in these regions, and therefore, the three-dimensional gravitational processes are needed to be employed. In the context of Newtonian gravity, \citet{Jog} explains that in a disc with a total thickness of $2h$,  the radial force in the mid-plane reduces by a factor of $(1 - e^{-kh})/kh $. Accordingly, the term $2 \pi G |\vec{k}| \Sigma$ in the dispersion relation modifies to $2 \pi G |\vec{k}| \Sigma(1 - e^{-kh})/kh $. To implement the role of the thickness in this modified dynamical model, one needs accurate formulation of potential of a thick disc which will be reported elsewhere \citep{ShenavarIII}.

The effect of the latter, i.e. the single-component assumption, however, seems to be more significant in star formation.  Some of the objects in our sample show a clear stellar-plus-gas effect, most notably one could name NGC 2976, NGC 2403. In the case of Newtonian model,   \citet{Wang} suggest that the Toomre's criterion of a stellar plus gas system is approximately $ Q_{sg} \simeq   \frac{\pi G}{\kappa} \left( \frac{\Sigma_{g}}{v_{s}} +\frac{\Sigma_{\star}}{\sigma_{r}} \right)^{-1}    $.
See also \citet{Romeo1} for another more accurate criterion which also reduces to \citet{Wang} in its proper limit. \citet{Wang} derive their result based on the method by \citet{Jog3}. In this method, they find the appropriate dispersion relation by treating the star-plus-gas system as two different isothermal fluids. The function $f(k)$ derived from dispersion relation has two distinctive peaks for the stellar and gas components. By expanding this function around these two peaks one could find the above modified $Q_{sg}$ for the stellar plus gas system.  However, \citet{Jog} has criticized this method because instead of finding a common two-component $k_{min}$, in which "min" corresponds to a minimum in $\omega^2$, \citet{Wang}  have  added the  contributions of the two systems at their respective neutral wavenumbers. The stability criterion of \cite{Wang} is also  independent of the gas fraction which is problematic. \cite{Jog}, on the other hand, presents another method to derive the local stability criterion of a two-component system. As one could easily see, finding  $Q_{sg}$ for the present model needs more careful investigations because the functional dependency of the fluid and stellar  dispersion relations to the wavenumber $k$ is more complicated. However, it is always possible to perform the necessary optimization numerically. This matter is beyond the present work.

\begin{figure}
\centering
\includegraphics[width=8cm]{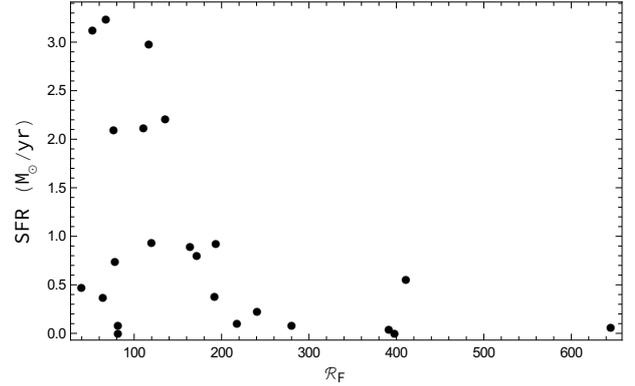}    \\
  \caption[]{Total SFR as a function of Freeman ratio $\mathcal{R}_{F}$ for different galaxies. As the plot suggests, HSB galaxies which posses low $\mathcal{R}_{F}$ show a higher SFR, if enough gas as the necessary component of star formation is provided, while LSBs with higher $\mathcal{R}_{F}$ show a lower SFR. The  present model predicts that LSBs are more stable than HSBs.}
  \label{fig:SFRvsRF}
\end{figure}

\begin{figure}
\centering
\includegraphics[width=8cm]{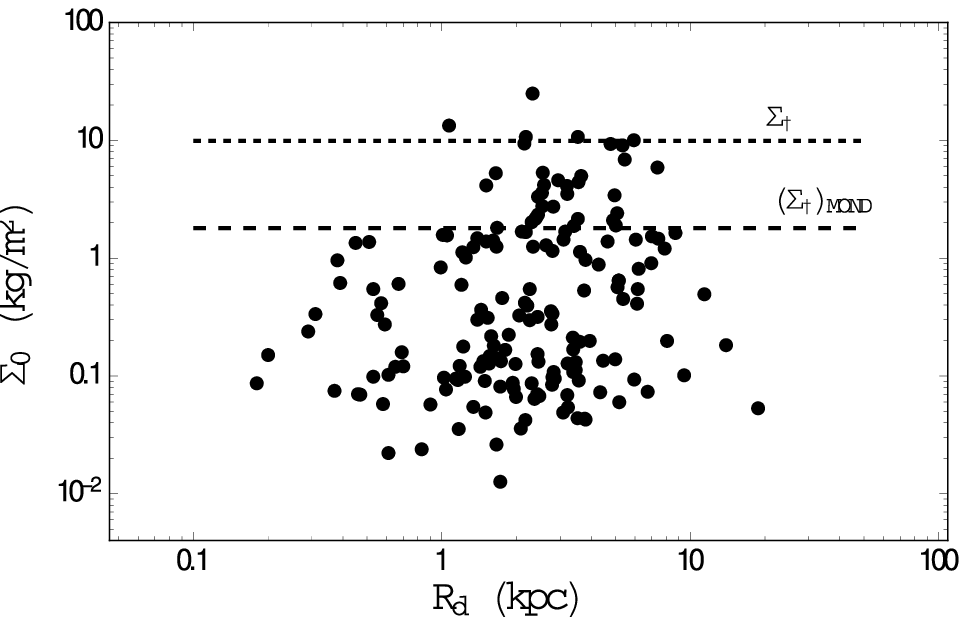}
\caption[]{ Surface density and size of  SPARC objects \citep{Sparc}.  This graph shows extrapolated $3.6~\mu m$ central surface density $\Sigma_{0}~~(kg/m^{2})$ of SPARC galaxies as a function of their  exponential  scale length of the stellar disc $R_d$. The number of galaxies declines when their $\Sigma_0$ exceeds $ \Sigma_{\dagger} \equiv a_0 /G $. The critical surface density of MOND, i.e. $(\Sigma_{\dagger})_{MOND}$, is almost 5.5 times smaller than $ \Sigma_{\dagger} $ because  we have $a_{0_{MOND}}=1.2 \times 10^{-10}m/s^{2}$ from  MOND phenomenology while here we have $a_{0}=cH_0 = 6.59 \times 10^{-10}m/s^{2}$ \citep{ShenavarII}.}
\label{FM}
\end{figure}

\textit{\textbf{ HSBs vs. LSBs:}}
Now we discuss the effect of Freeman ratio $\mathcal{R}_{F}$ on the stability. Observations show that the central surface density of a large number of Low Surface Brightness(LSB) galaxies is about one order of magnitude lower than their High Surface Brightness (HSB) counterparts \citep{schom, Mihos}. Although LSBs have a normal amount of HI \citep{Impey},  their star formation rate is low and this phenomenon have been attributed to the lack of dust and molecular gas \citep{kruitFree}. In the CDM model, the LSBs are believed to be dominated by dark matter haloes from the innermost radii \citep{Ghosh} and observations show that they are mostly stable against global non-axisymmetric instabilities \citep{Mihos}. In a recent work, \citet{Ghosh} by using a global  analysis within WKB approximation have studied the DM influence on collisionless galactic discs and found compatible results with observations. Despite the low amount of surface density, the scarcity of star formation in LSB discs is also suggested to be an effect of their local environment, since they are usually seen to be less clustered, and also mostly isolated on the scales less than a few Mpc. However, if they are influenced by tidal effects in denser environments, they can be transformed to HSB galaxies or even be destroyed entirely. Otherwise they are believed to be only passively evolved \citep[and references therein]{Mihos}.

In the present work, we argued in the previous sections  that LSB galaxies are generally more stable than HSBs and so should show less star formation. See Tab. \ref{table:QqQs} for a comparison. For the THINGS sample, we have plotted total SFR of the objects versus $\mathcal{R}_{F}$ in Fig. \ref{fig:SFRvsRF}. As this plot shows, galaxies with a high value of $\mathcal{R}_{F}$, i.e. LSBs, show a lower SFR. On the other hand, those with relatively lower $\mathcal{R}_{F}$, i.e. HSBs, display high SFR. Of course, the star formation could occur in HSBs if there is enough gas in the environment; thus, we see that there are some HSBs with in fact low SFR. These objects have supposedly consumed most of their gas reserve in the past to form stars. Admittedly though, one needs a much larger sample to draw a more definite conclusion.

\textit{\textbf{  The role of $\Sigma_{\dagger}$ scale in stability:}} Now we check the role of $\Sigma_{\dagger}$ in the stability of galactic discs, as mentioned in Sec. \ref{MD} above.  Plot \ref{FM} shows the surface density and size of the SPARC sample of \cite{Sparc}. The SPARC sample includes 175 galaxies with a broad range of morphologies. The extensive variety of this sample helps to examine the role of  $\Sigma_{\dagger}$ more accurately. Similar to \cite{Sparc},  we have assumed a mass-to-light ratio as $ \Upsilon_{\star} =0.5 M_{\odot}/L_{\odot}$.  As you could see from Fig. \ref{FM}, the  central surface density of galaxies rarely exceeds  $ \Sigma_{\dagger} $ which is expected in our model because, as we discussed above, discs with central surface densities close to $ \Sigma_{\dagger} $ are more unstable compared to cases with $ \Sigma \ll \Sigma_{\dagger} $. In fact, priorly there have been some observational reports which suggest the existence of an upper limit on the  surface density of galactic discs. Most notably, \cite{McGaugh96} has surveyed disc galaxies and found that although HSB galaxies are easier to observe compared to LSBs, because the former type has higher surface density than the latter, the number of HSB galaxies with surface densities exceeding $\Sigma_{\dagger}$ decreases exponentially. Such behaviour has been predicted  by \cite{Milgrom5} for the first time based on stability of discs in MOND model. See also Figure 8 in \cite{mond}, though, one should remember that in this paper the authors use $(\Sigma_{\dagger})_{MOND}$ which is almost 5.5 times smaller than $\Sigma_{\dagger}$ here. The critical surface density of MOND is also shown in Fig. \ref{FM} for comparison. On the other hand,  in Newtonian gravity, dark matter haloes  could stabilize  disc galaxies with very high -- much higher than observed -- surface densities \cite{mond}. To see this, consider a galactic disc surrounded by a dark matter halo.  Assuming that such a configuration with a certain dynamics is stable, if one multiplies the  density by a positive constant and  scales the particles' velocities appropriately, it is essentially possible to build another stable system. As \cite{mond} have explained, the absence of a clear mechanism to provide an upper limit on  central surface density has motivated \cite{Spergel} to introduce it into the models by hand. 

\textit{\textbf{  Improving the predictions of model II:}}
A more accurate investigation of the two models shows that  their differences are mostly based on distinct behaviours of  their epicyclic frequency $\kappa$, which in turn is derived from different rotation curve formulas.  The epicyclic frequency of the present modified dynamical model is based on Eq. \eqref{Rcurve1}.  The   predictions of model II, on the other hand, are based on  $v_{Boiss}$ as the governing law of RC  which is not very accurate at small radii.   In fact, \citet{Leroy} admit that small-scale details of  RC  are lost in their data fittings  presumably due to streaming motions near spiral arms or warps in the gas disc. Small radii parts of the galaxies are  usually regions with high observed $\Sigma_{SFR}$. This is exactly the source of the problematic behaviour of the model II.    We saw above that the prediction derived from Eq. \eqref{Boissier} for THINGS catalog displays a very high value of epicyclic frequency toward the center. See Fig. \ref{fig:finalplt} for more details. This behaviour essentially leads to a high critical surface density   for Newtonian model, because  $\Sigma^{crit} \propto \kappa  $,  which ultimately predicts a low ratio of $\Sigma/\Sigma^{crit}$ for both gas and stellar component. Of course, one could re-estimate the rotation curve parameters $v_{flat}$ and $l_{flat}$ so that the \citet{Boissier} proposal becomes more plausible at small radii, but then it might show problematic behaviour at larger scales. On the other hand, if one uses other rotation curve formulae, say for example  \citet{Brandt} law, the predictions will certainly change. Unfortunately, for the THINGS catalogue, only the properties based on Eq. \eqref{Boissier} are reported by \citet{Leroy} and so we are not able to test other models of  rotation curves in this work. 

 Instead of relying on observationally-based RC formulas like $v_{Boiss}$, one could check the local stability of the Newtonian model based on different dark halo profiles. For example,  \citet{Jog2}  derives the modified $\kappa$ parameter for a disc in a spherical  pseudo-isothermal   dark matter halo. As discussed above, this modified epicyclic frequency could be found as $ (\kappa_{disc}^{2}+\kappa_{halo}^{2})^{1/2}$ in which $ \kappa_{halo} $ adds to the stability of the disc. \citet{Jog2} results will not be used here because the rotation curve parameters related to pseudo-isothermal   dark matter halo are not reported by \citet{Leroy}.

 However, one should note that although the dark matter halo has a stabilizing effect through increasing $\kappa$, it also destabilizes the system by increasing the local surface density  $ \Sigma_{d0} $ which now consists of $ \Sigma_{halo}$ too \citep{Corbelli}. This could be simply seen from the definition of Toomre parameter $Q_{g} = \kappa v_{s} / (\pi G \Sigma_{d0}) $. One might argue that the share of dark particles in local surface density is negligible; however,  \citet{Weber} have studied various halo profiles of the Galaxy and have shown that while the surface density of baryonic matter in local neighbourhood is around $48 \pm 9~~M_{\odot}/pc^{2}$, the total surface density is about  $71 \pm 6~~M_{\odot}/pc^{2}$. Thus, the dark matter particles contribute almost $30 \%$ to the local surface density which is not dominant but still considerable. The combined effect of $\kappa_{halo}$ plus  $\Sigma_{halo}$ needs more careful studies for different halo profiles and is  beyond the scopes of the present work.

 \citet{deBlok}  have analysed rotation curves of  dwarf and LSB galaxies by using  core dominated pseudo-isothermal and also  cusp dominated haloes. Their conclusion is that for these objects the pseudo-isothermal profile provides better fits to  the data compared to NFW profile because the latter results in very low $M/L$ ratios.  NFW profile is derived based on numerical  simulations of cold dark matter model \citep{NFW1,NFW2}. From "gravitational stability" point of view, a comparison between pseudo-isothermal and NFW profiles seems to be quite interesting. In this respect, one could contrast the behaviours of $\Sigma_{SFR}$ of the two profiles for  galaxies with different dynamics and morphologies.

\section{\small{Conclusion}} 
\label{conc}
In this work, we considered the local stability of a modified dynamical model which is based on imposing Neumann BC on GR field equations. Both fluid and stellar discs in this model are found to be locally more stable compared to pure Newtonian discs. For the gaseous (stellar) discs, when  $\beta_{g} $ ( $\beta_{\star} $) increases the maximum value that is needed for $Q$ to make the system stable decreases. Thus, regions with larger $\beta_{g} $ or $\beta_{\star} $ are more stable. Also, LSB galaxies which show a high value of  Freeman ratio $ \mathcal{R}_{F} \equiv \frac{\Sigma_{\dagger}}{ \Sigma_{0}}$ are predicted to be more stable than HSBs. The growth rate were found to be smaller than Newtonian model for both fluid and stellar discs. We tested the model by using THINGS data of \citep{Leroy} and showed that the general behaviour  of  star formation rate per surface density ($\Sigma_{SFR}$) agrees  with  the predictions of the model.

Other modified theories too have tried to explain the issue of local stability and many interesting results have been reported. For example,  \citet{Milgrom5} reports that MONDian discs are more stable than Newtonian ones. Also, in the deep MOND limit the stability criteria becomes independent of the acceleration. Furthermore, \citet{Milgrom5} argues that when $\Sigma \gg \Sigma_{\dagger}$, or equivalently when $\mathcal{R}_{F} $ is  small, the Newtonian regime prevails and the disc  becomes  unstable. It is then proposed  that the rarity of discs with small $ \mathcal{R}_{F}$ is due to their unstable nature.  Using \citet{Sparc} data, we observed the same behaviour here as shown in Fig. \ref{FM}. Also, Fig. \ref{FM} provided a chance to compare the role of $\Sigma_{\dagger}$ in MOND and the present model.

\citet{Roshan1,Roshan2} show that although MOG provides more attracting force compared to Newtonian theory, it is interestingly less stable than Newtonian case. However, they argue that the difference between the two theories is too small to be detected at galactic scale. This is shown by introducing a parameter $\beta_{MOG} \equiv \mu_{0}v/\kappa$, counterpart to $\beta $ in this work, in which $\mu_{0} =0.042 \pm 0.004 ~ kpc^{-1}$ is a fundamental constant of MOG derived by \citet{Rahvar}. Any difference between the stability of MOG and Newtonian models is proportional to $ \beta_{MOG} $, and it would be negligible because of the smallness of $ \mu_{0} $ at galactic scales. If, however, this parameter is dependent to the system, say for instance $ \mu_{0} \propto 1/R_{d} $ in analogy with our analysis, then the instability of MOG could be studied at galactic scales. In fact, \citet{Haghi} suggest that to fit MOG to the observed velocity dispersion of  dwarf spheroidal galaxies, one needs to consider  $ \mu_{0}$ as a varying parameter. This possibility, however, complicates the MOG model even more. In the case of $f(R)$ theories, on the other hand, \citet{Roshan3} show that the discs in a specific class of  these theories  are generally more stable than Newtonian discs.

Another challenge facing modified gravity/dynamics models is the problem of global stability. For example, \citet{Roshan4} develop a semi-analytic method to study the response of a Maclaurin disc to linear non-axisymmetric perturbations for the theory of MOG. Their results show that  Maclaurin discs are less stable in MOG compared to Newtonian gravity and especially the bar mode, i.e. $m=2$, is strongly unstable and unlike in Newtonian gravity cannot be avoided. Also \citet{Neda} study the  global stability of Mestel and exponential  discs in MOG from numerical point of view. In future works, we will examine the global stability of the present model in analytical and numerical studies. The equilibrium configuration of our model is reported in \citet{ShenavarIII}.

\section*{Acknowledgements}
This work is partially supported by the Iranian Ministry of Science, Research and Technology (MSRT) and Iran Science Elites Federation. We  feel grateful to Pavel Kroupa  for his useful comments and also his support during our visit at Bonn University. We would like to thank  Mahmood Roshan for various discussions.  We also thank the authors Leroy,  Walter, Brinks, Bigiel,  de Blok, Madore,  Thornley (THINGS) and Lelli,  McGaugh,  Schombert (SPARC) for sharing their data freely. We acknowledge the anonymous reviewer whose comments helped to clarify and strengthen the discussions and presentation of the work.   This research has made use of NASA’s Astrophysics Data System.



\bibliographystyle{mnras}
\bibliography{local1} 







\bsp	
\clearpage

\begin{figure*}{h!}
\centering
\includegraphics[height=19cm,width=15cm]{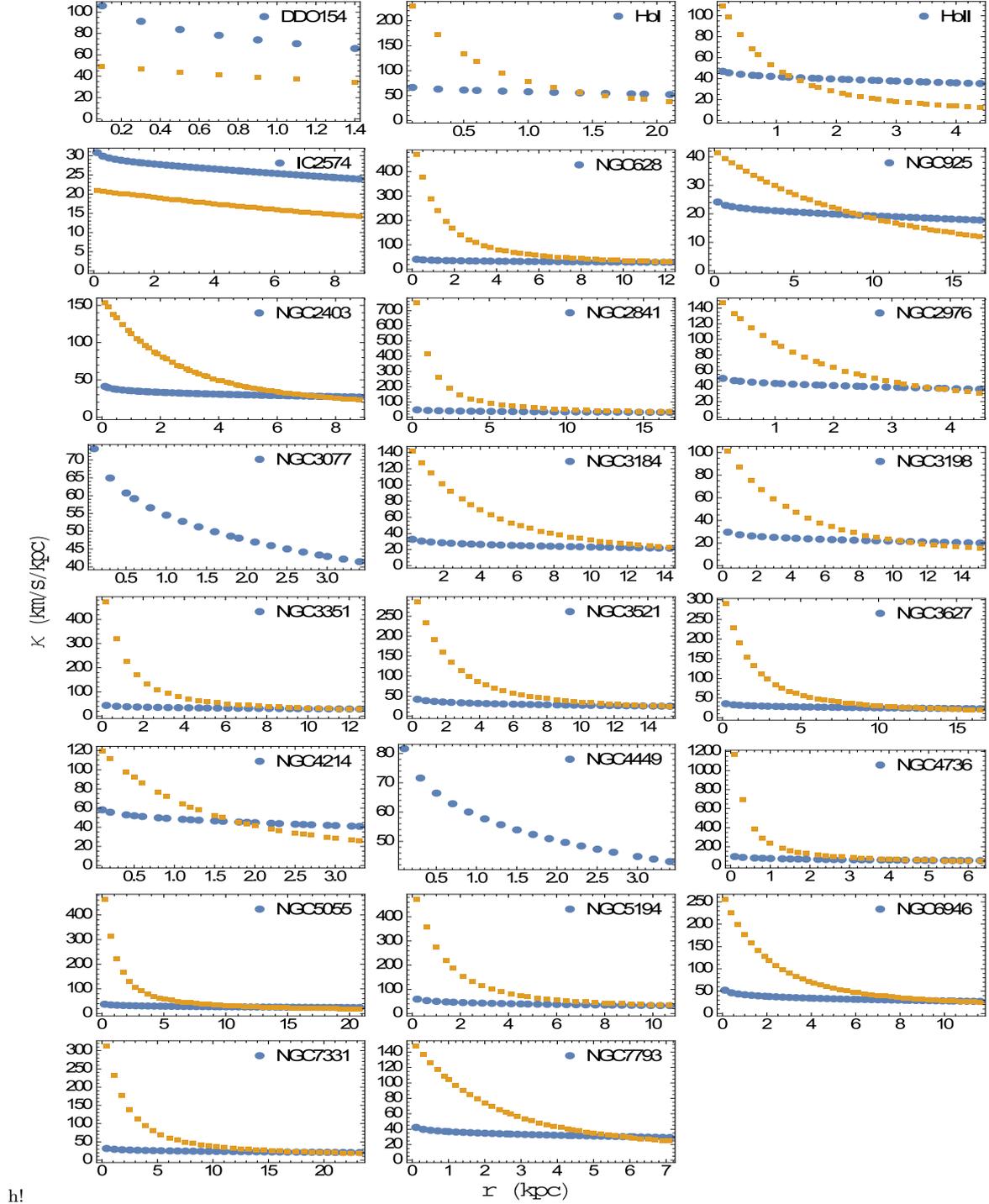}    \\
  \caption{Epicyclic frequency $\kappa $ of  the present model, shown by blue points, and  model II which is shown by yellow squares.    As it is clear from this graph, in the interior parts of most galaxies, $\kappa $ from model II is typically much larger than $\kappa_{MOD} $ of the present model. This leads to a higher critical surface density for model II, compared to our model, which in turn damps the ratio $\Sigma/\Sigma_{crit}$ that is related to $\Sigma_{SFR}$. }
  \label{fig:kappa}
\end{figure*} 
\pagebreak

\begin{figure*}
\centering
\includegraphics[height=22cm,width=17cm]{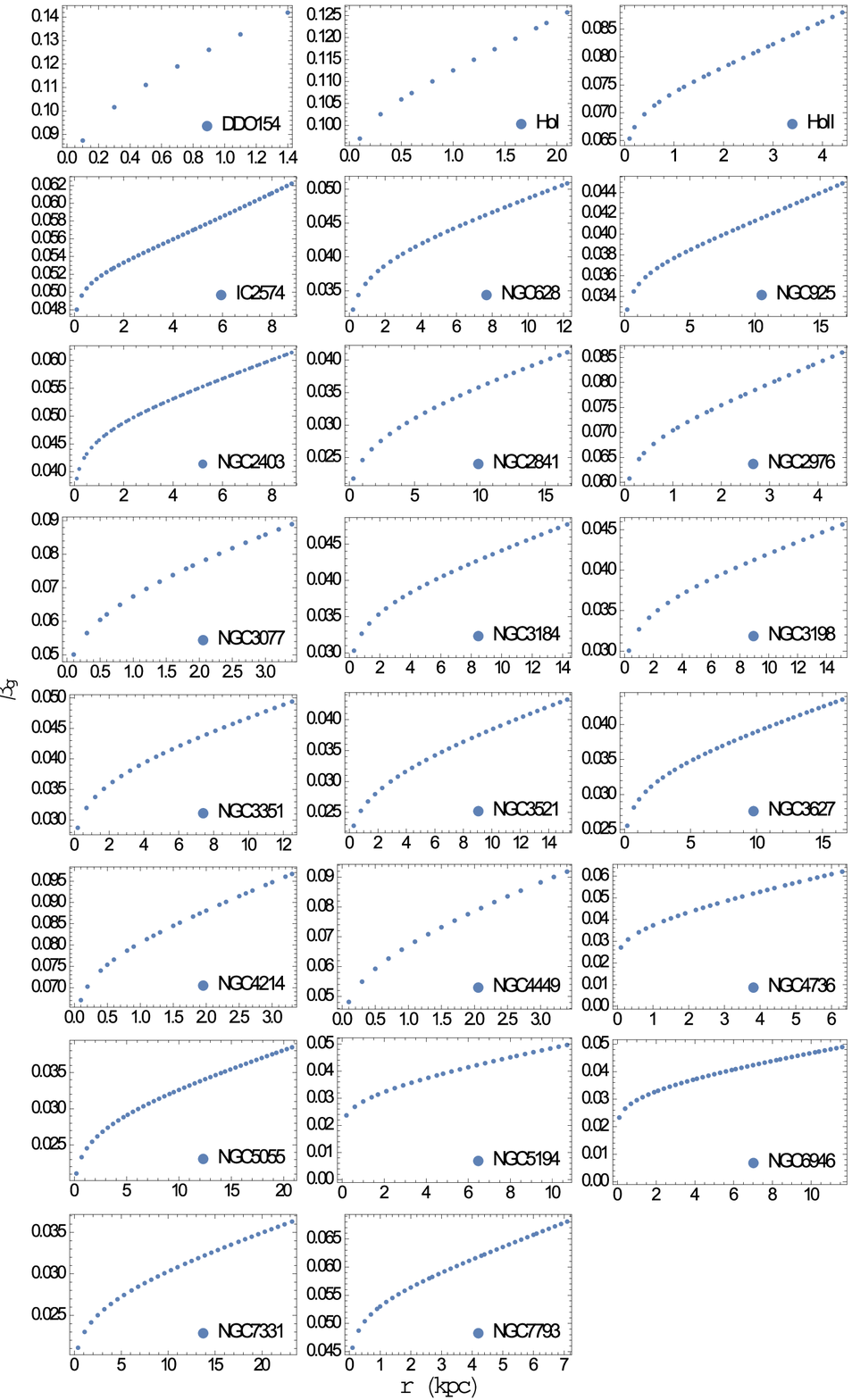}    \\
  \caption{$\beta_{g} $ as a function of radius r (kpc) for the stellar disks.  }
  \label{fig:betastellar}
\end{figure*}
\pagebreak

\begin{figure*}
\centering
\includegraphics[height=22cm,width=17cm]{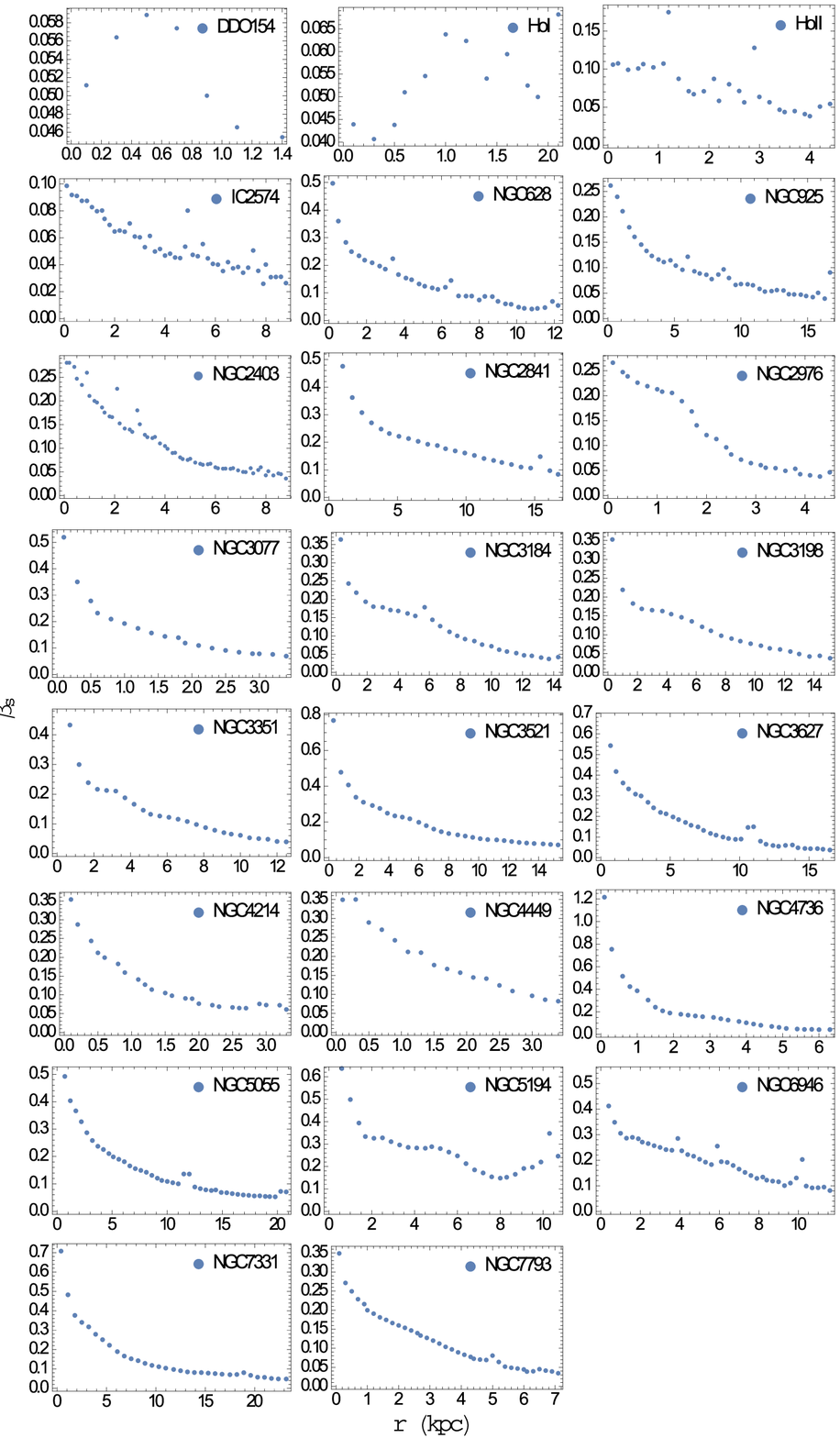}    \\
  \caption{$\beta_{\star} $ as a function of radius r (kpc) for the stellar disks.  }
  \label{fig:betastellar}
\end{figure*}
\pagebreak

\begin{figure*}
\centering
\caption{The First subsample. Left panels are $\Sigma_{\star}/\Sigma^{crit}_{\star} $, the middle panels are $\Sigma_{gas}/\Sigma^{crit}_{gas} $ while the panels on the right are total star formation rate surface density  $\Sigma_{SFR}$  in units of $10^{-4}~ M_{\odot}/yr/kpc^{2}$.  Data  from Leroy et al. (2008). In the left and middle panels the results of the present model are shown by  points (blue) while predictions of model II  are shown by yellow squares (colour online). In this subsample, the star formation rate surface density mostly follows the pattern of $\Sigma_{gas}/\Sigma^{crit}_{gas} $.}
\includegraphics[height=20cm,width=15cm]{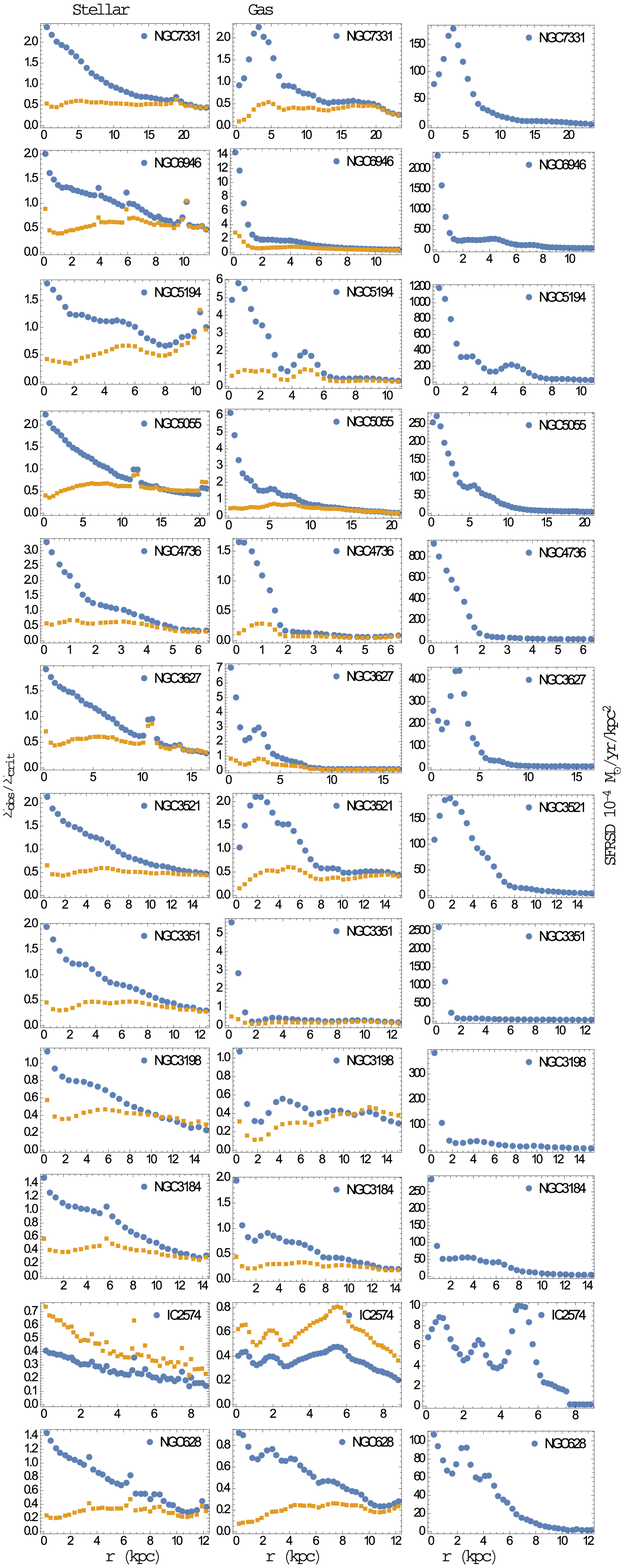}   
\end{figure*}
\pagebreak

\begin{figure*}
\centering
\includegraphics[height=19cm,width=15cm]{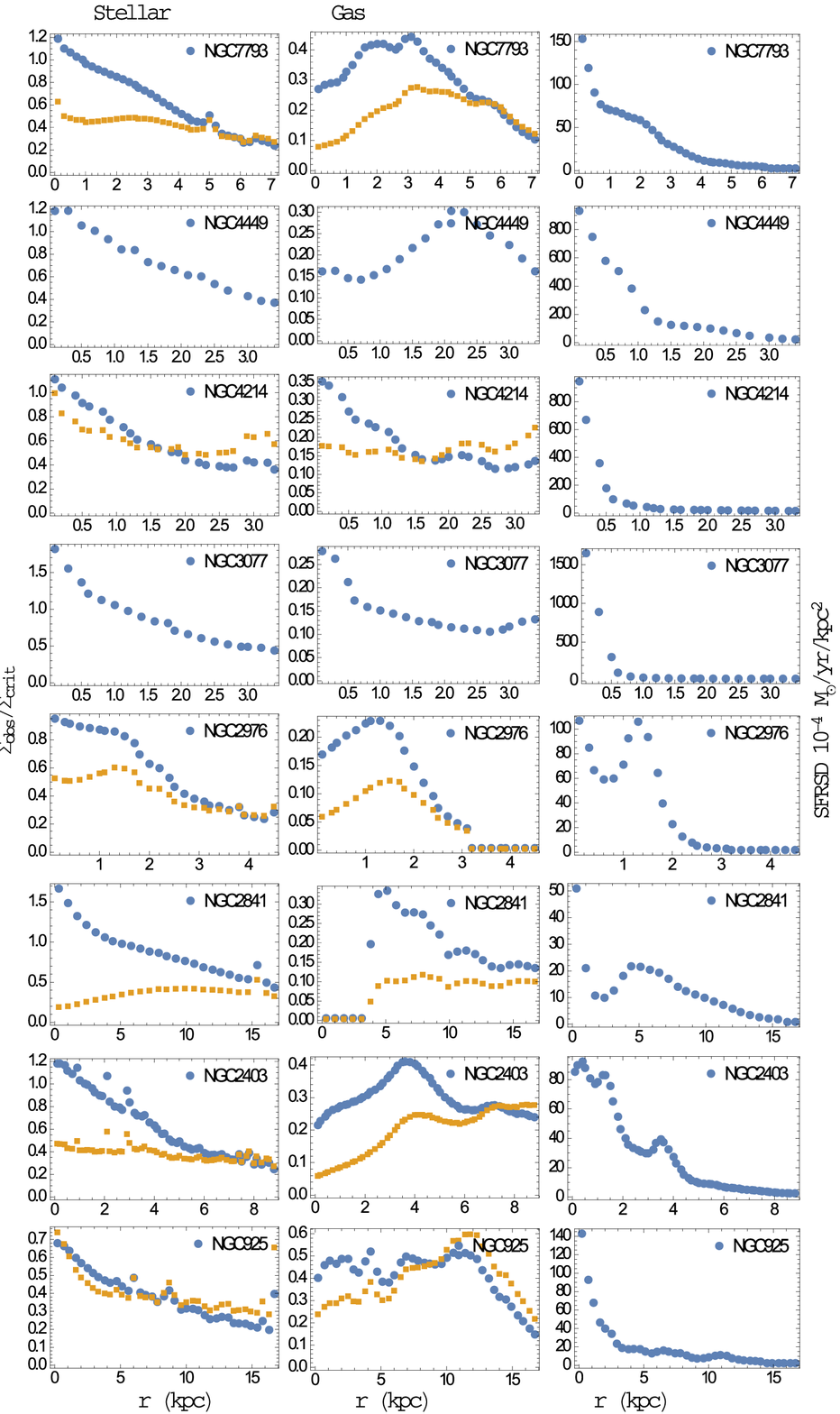}   
  \caption{The second subsample. Left panels are $\Sigma_{\star}/\Sigma^{crit}_{\star} $, the middle panels are $\Sigma_{gas}/\Sigma^{crit}_{gas} $ while the panels on the right are total star formation rate surface density  $\Sigma_{SFR}$  in units of $10^{-4} M_{\odot}/yr/kpc2$.  Data  from Leroy et al. (2008). In the left and middle panels the results of the present model are shown by blue points  while predictions of model II  are shown by yellow squares (colour online). In this subsample, the patterns in  $\Sigma_{\star}/\Sigma^{crit}_{\star} $ and  star formation rate surface density  $\Sigma_{SFR}$ are qualitatively similar.}
\end{figure*}
\pagebreak

\begin{figure*}
\centering
\includegraphics[width=15cm]{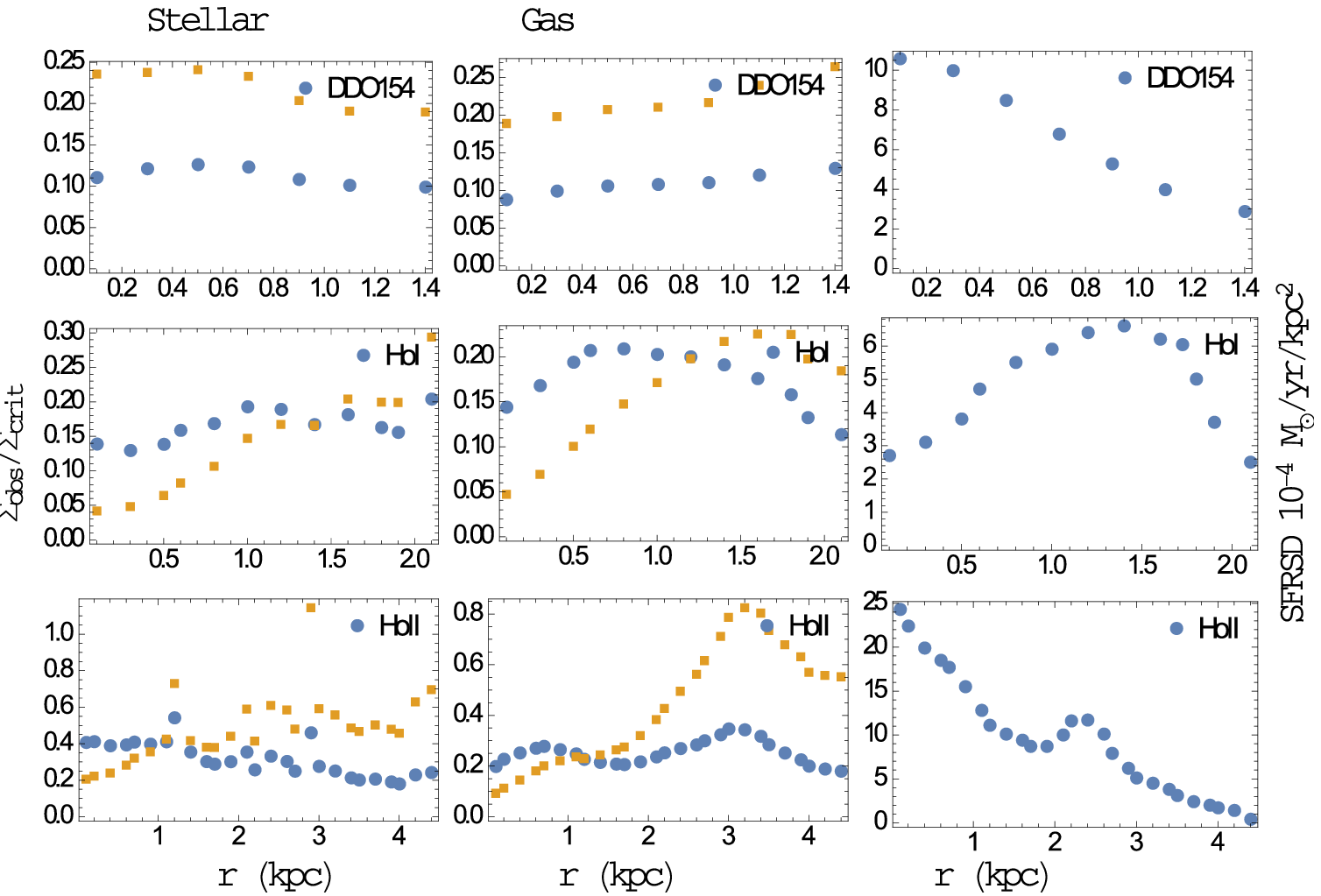}   
  \caption{The third subsample. Left panels are $\Sigma_{\star}/\Sigma^{crit}_{\star} $, the middle panels are $\Sigma_{gas}/\Sigma^{crit}_{gas} $ while the panels on the right are total star formation rate surface density  $\Sigma_{SFR}$  in units of $10^{-4} M_{\odot}/yr/kpc2$.  Data  from Leroy et al. (2008). In the left and middle panels the results of the present model are shown by  points (blue) while predictions of model II  are shown by yellow squares (colour online). For these three irregular galaxies, there is no obvious similarity between $\Sigma_{SFR}$ and  $\Sigma_{\star}/\Sigma^{crit}_{\star} $ or $\Sigma_{gas}/\Sigma^{crit}_{gas} $.}
\end{figure*}


\label{lastpage}
\end{document}